\documentclass{ieeeaccess}
\usepackage{cite}
\usepackage{amsmath,amssymb,amsfonts}
\usepackage{algorithmic}
\usepackage{graphicx}
\usepackage{textcomp}
\usepackage{array}
\usepackage[caption=false,font=normalsize,labelfont=sf,textfont=sf]{subfig}
\usepackage{url}
\usepackage{verbatim}
\usepackage{cite}
\usepackage{booktabs}
\usepackage{tablefootnote}
\usepackage{threeparttable}
\usepackage{xcolor}
\usepackage{soul}
\newcommand{\comm}[1]{}
\usepackage{bm}

\makeatletter
\AtBeginDocument{\DeclareMathVersion{bold}
\SetSymbolFont{operators}{bold}{T1}{times}{b}{n}
\SetSymbolFont{NewLetters}{bold}{T1}{times}{b}{it}
\SetMathAlphabet{\mathrm}{bold}{T1}{times}{b}{n}
\SetMathAlphabet{\mathit}{bold}{T1}{times}{b}{it}
\SetMathAlphabet{\mathbf}{bold}{T1}{times}{b}{n}
\SetMathAlphabet{\mathtt}{bold}{OT1}{pcr}{b}{n}
\SetSymbolFont{symbols}{bold}{OMS}{cmsy}{b}{n}
\renewcommand\boldmath{\@nomath\boldmath\mathversion{bold}}}
\makeatother

\def\BibTeX{{\rm B\kern-.05em{\sc i\kern-.025em b}\kern-.08em
    T\kern-.1667em\lower.7ex\hbox{E}\kern-.125emX}}

\begin{document}
\history{Date of publication xxxx 00, 0000, date of current version xxxx 00, 0000.}
\doi{10.1109/ACCESS.2024.0429000}

\title{Fatigue-PINN: Physics-Informed Fatigue-Driven Motion Modulation and Synthesis}
\author{\uppercase{Iliana Loi}\authorrefmark{1},
\uppercase{Konstantinos Moustakas}\authorrefmark{2}
~\IEEEmembership{Senior Member,~IEEE}}

\address[1]{Department
of Electrical and Computer Engineering, University of Patras, Patras, 26504 Greece (e-mail: loi@ceid.upatras.gr)}
\address[2]{Department
of Electrical and Computer Engineering, University of Patras, Patras, 26504 Greece (e-mail: moustakas@ece.upatras.gr)}
\tfootnote{The data used in this project was obtained from \url{https://fling.seas.upenn.edu/~mocap/cgi-bin/Database.php}. The research project is implemented in the framework of H.F.R.I call “Basic research Financing (Horizontal support of all Sciences)” under the National Recovery and Resilience Plan “Greece 2.0” funded by the European Union -- NextGenerationEU (H.F.R.I. Project Number: 16469.).}

\markboth
{Loi \headeretal: Fatigue-PINN: Physics-Informed Fatigue-Driven Motion Modulation and Synthesis}
{Loi \headeretal: Fatigue-PINN: Physics-Informed Fatigue-Driven Motion Modulation and Synthesis}

\corresp{Corresponding author: Iliana Loi (e-mail: loi@ceid.upatras.gr).}

\begin{abstract}
Fatigue modeling is essential for motion synthesis tasks to model human motions under fatigued conditions and biomechanical engineering applications, such as investigating the variations in movement patterns and posture due to fatigue, defining injury risk mitigation and prevention strategies, formulating fatigue minimization schemes, and creating improved ergonomic designs. Nevertheless, employing data-driven methods for synthesizing the impact of fatigue on motion, receives little to no attention in the literature. In this work, we present \textit{Fatigue-PINN}, a deep learning framework based on Physics-Informed Neural Networks, for modeling fatigued human movements, while providing joint-specific fatigue configurations for adaptation and mitigation of motion artifacts on a joint level, resulting in \comm{more realistic}more smooth, hence physically-plausible animations. To account for muscle fatigue, we simulate the fatigue-induced fluctuations in the maximum exerted joint torques by leveraging a PINN adaptation of the Three-Compartment Controller model to exploit physics-domain knowledge for improving accuracy. This model also introduces parametric motion alignment with respect to joint-specific fatigue, hence avoiding sharp frame transitions. Our results indicate that Fatigue-PINN accurately simulates the effects of externally perceived fatigue on open-type human movements being consistent with findings from real-world experimental fatigue studies. Since fatigue is incorporated in torque space, Fatigue-PINN provides an end-to-end encoder-decoder-like architecture, to ensure transforming joint angles to joint torques and vice-versa, thus, being compatible with motion synthesis frameworks operating on joint angles.
\end{abstract}


\titlepgskip=-21pt

\maketitle

\section{Introduction}
\label{sec:introduction}
\PARstart{S}{tuding} the effects of fatigue in human motion has been the focus of many biomechanical works \cite{Cowley2017, Yang2019, Haralabidis2020, He2024}, where the investigation of possible variations in movement patterns and posture and the inevitable decrease in maximum exerted torques and muscle forces as well as changes in the range of joint angles under fatigue conditions, is conducted. 
In contrast to typical biomechanics research, such as the previously mentioned studies that do not incorporate modeling of fatigued motions, data-driven approaches for estimating musculoskeletal dynamics (e.g. \cite{Sharma2022, Mansour2023}) offer promising and automated solutions for ergonomically-adjusted motion estimation. This includes fatigue-driven motion modeling, which is rather difficult to achieve without knowing the internal human state.
Especially, joint torques are inextricably correlated with muscle activity and can be defined as a function of joint angles \cite{Anderson2007}. Thus, the determination of joint torques is based on joint angles, velocities, and accelerations, i.e. Inverse Dynamics (ID), and vice versa, i.e. Forward Dynamics (FD). 
Machine and Deep learning approaches (ML/DL) for estimating physical movement parameters, such as joint contact forces (particularly those of the knee) \cite{Loi2023, Zou2024}, joint torques \cite{Wang2023, Mansour2023}, and muscle forces \cite{Sohane2022}, function as surrogate models for Inverse and Forward dynamics. These approaches can significantly enhance the simulation of physically-plausible 3D humanoid character movements.
In recent years, state-of-the-art approaches such as Physics-Informed Neural Networks (PINNs)\comm{, which embed physics-based domain knowledge into the training process,} have been employed to bridge the gap between kinematics and dynamics estimation techniques \cite{Zhang2023, Ma2024, ZhiboZhang2022}.

However, fatigued-driven human motion generation using data-driven methods is yet to be explored despite its various advantages, like i) risk mitigation and prevention of injuries by understanding how fatigue affects movement patterns in both high-performance athletic \cite{He2024} and non-athletic physically demanding occupations \cite{Jo2024}, ii) design of improved rehabilitation protocols by tailoring exercises that consider the impact of fatigue on muscle performance and joint stability \cite{Carratalá-Tejada2022}, iii) optimized performance in athletes \cite{Smits2014, Dunn2017, Dunn2019} and workers \cite{Lu2021} by using strategies that minimize the negative effects of physical fatigue, 
iv) creation of better ergonomic designs (e.g.equipment, tools, workspaces, etc.) to reduce fatigue and enhance comfort and productivity \cite{Sharotry2022}, as well as v) the production of realistic simulations and physically-plausible fatigued-driven animation without the recording of strenuous motion capture (gaming) or handcrafted animation sequences (animation), which is essential for applications in virtual reality, gaming, ergonomic assessments, and rehabilitation. 

To address this gap in the literature, we introduce a deep learning framework based on PINNs to produce fatigued open-type human movements. As open-type human motion, we refer to both non-periodical movements and motions not involving human-object interactions. To account for muscle fatigue, we simulate the fatigue-induced fluctuations modulating joint torques of a 3D humanoid character. This simulation is grounded in a PINN adaptation of the Three-Compartment-Controller (3CC) Model \cite{Liu2002, Xia2008, Frey-Law2012}, \textit{3CC-$\lambda$}, a state machine that describes the transition of all muscle motor units of a human limb from one state (compartment) to another, namely active ($M_A$), fatigued ($M_F$), or resting ($M_R$) states. Given that the majority of motion synthesis frameworks produce sequences of joint angles, we devised an end-to-end encoder-decoder-like architecture, namely \textit{Fatigue-PINN}, to incorporate fatigue effects within the torque space. In this setup, a surrogate DL model for ID converts joint angles into joint torques, then, the 3CC-$\lambda$ model introduces fatigue to generate fatigued torques, which are afterward processed by a surrogate DL model for FD to perform the reverse transformation. Our semi-dynamics approach is joint-specific, meaning that Fatigue-PINN, is applied for estimating fatigue in each joint, ensuring the generation of \comm{realistic}smooth fatigue-driven animation without explicitly modeling the complex underlying physics that governs the human body and without the acquisition of fatigued motion capture data for training our models. It is worth noting that fatigued motion capture datasets containing open-type actions or more sophisticated movements, which can be used for training fatigued motion synthesis DL models, are absent from the literature. The latter, along with the lack of data-driven models for the generation of fatigued human motion, serve as the motivation behind this work.
Our contributions are summarized as follows:
\begin{itemize}
    \item Introduction of Fatigue-PINN, an end-to-end automated PINN-based framework to model fatigue in open human movements by exploiting internal human state domain knowledge, bypassing the necessity of acquiring fatigued motion capture data. Our model provides joint-specific fatigue configurations allowing adaptation and mitigation of motion artifacts in a semi-automated manner, resulting in \comm{more realistic}smoother animations. 
    \item Using 3CC-$\lambda$, a PINN variation of 3CC, to model fatigue in an automated and physically-consistent manner. \comm{PINNs exploit physics-based domain knowledge to constraint the estimation of the output, hence accelerating convergence and accuracy in cases of few training data.}
    \item Provide robust deep learning methods to pose as semi-dynamics surrogate models for ID and FD procedures without taking into account Ground Reaction Forces (GRFs).
    \item The end-to-end encoder-decoder-like architecture of Fatigue-PINN enables its seamless integration into any animation pipeline that operates on joint angles. 
\end{itemize}

A demo of Fatigue-PINN is made available at GitHub\footnote{\url{https://github.com/loi10/Fatigue-PINN}}.

\section{Related Work}
\subsection{Data-Driven Motion Synthesis}
 
In the motion synthesis area, researchers and developers in the graphics, animation, and gaming industries, dive deeply into both deterministic and probabilistic Deep Learning (DL) motion generation
techniques. These data-driven approaches leverage human motion capture and movement history data (e.g., previous frames or motion states like joint angles) to produce or predict the pose and/or joint trajectories of a 3D humanoid character \cite{Loi2023_review}. Especially for deterministic motion synthesis, where the synthesized motion sequence converges to a deterministic pose sequence that regresses towards the mean pose of ground-truth, recurrent architectures like LSTM models \cite{Aristidou2021} for typical motion imitation and RNNs enhanced with phase-functioned networks \cite{Starke2019, Starke2020, Starke2021} for scene-aware interactions, are favored. The work in \cite{Starke2019} introduces the Neural State Machine (NSM), a framework for goal-driven real-time synthesis of 3D character movements and scene interactions. The NSM consists of a gating network and an RNN motion prediction network, where the first accounts for automatic action state transition based on the global phase of the motion signal and the user input (i.e. desired goal). In the subsequent works of the same authors \cite{Starke2020, Starke2021}, an architecture similar to NSM is employed and enhanced with: i) a local phase feature that exploits the local motion phases of each skeletal segment to dynamically produce their asynchronous motion in character-character, character-object, and character-scene interactions in \cite{Starke2020}, and ii) a control scheme in \cite{Starke2021} for animation layering to produce sophisticated (e.g. martial arts movements) and novel motions based on reference motions, physics-based simulations, input controls, etc. Convolutional Neural Networks (CNNs) have also been employed for synthesizing human motion sequences \cite{Zhou2019}.

On the contrary, Generative Adversarial Networks (GANs) \cite{Men2021, Mourot2022}, Variational Autoencoders \cite{Cai2021, Zhou2023}, Transformers \cite{Hou2024, Chai2024}, and the state-of-the-art diffusion models for motion generation conditioned on multiple sources (e.g. natural language descriptions and audio) \cite{Raab2023, Alexanderson2023, Dabral2023, Gao2024}, are exploited for probabilistic motion synthesis meaning the construction of all plausible pose sequences of a virtual character based on historical poses and/or control inputs. Probabilistic DL models have the ability to inject stochasticity in the training process, i.e. by fitting a latent distribution to the distribution of the next pose in each invocation, and, hence, increasing the learning capacity of the model.

Motion synthesis works can be used alongside musculoskeletal dynamics estimation approaches (described in Section \ref{sec:Musculoskeletal}) to reflect the human internal state on a 3D humanoid character, thus, providing physics-based solutions. Consequently, works in the physics-based motion synthesis area develop holistic approaches, usually relying on Reinforcement Learning (RL) \cite{Lee2021, Luo2024}, which account for the data-driven imitation of physically realistic motion sequences conditioned on physical parameters such as joint velocities, forces, torques, etc. It is worth mentioning the most recent work in \cite{Zhang2024}, where an auto-regressive neural network resembling previous work \cite{Starke2019, Starke2020, Starke2021}, which incorporates knowledge from the physical attributes of exerted human force and perceived resistance, was applied to model variations in human movements while interacting with objects and, thus, enhancing realism.

Nevertheless, even though great steps are made towards more realistic computer graphics character animation, all the aforementioned methods and the ones referenced in Section \ref{sec:Musculoskeletal}, focus on modeling the active state of human motion. Fatigued motion using DL has received little to no attention in the literature. To the best of our knowledge, only one work, the one in \cite{Cheema2023}, attempts fatigue-driven animation, using RL methods for controlled motion imitation, and the Three-Compartment Controller (3CC) state machine to model torque-based fatigue. 

\subsection{Data-Driven Musculoskeletal Dynamics and Kinematics Estimation}
\label{sec:Musculoskeletal}

Estimating musculoskeletal human biomechanics has been the focus of research for many years, with results affecting multiple research fields such as biomechanics, the gaming industry, robotics, and the animation industry. Biomechanics engineers use powerful open-source tools like OpenSim \cite{Seth2018}, to simulate with great precision human kinematics and compute the kinetics and dynamics of a motion (e.g. joint contact forces, joint torques, muscle forces, etc.), while state-of-the-art methods focus in data-driven approaches (ML/DL), such as the ones in \cite{Sharma2022, Loi2023, Wang2023, Mansour2023} to obtain more automated, faster as well as real-time, solutions while estimating biomechanical variables traditionally calculated through musculoskeletal modeling. Such works rely on raw motion marker data, joint kinematics data (i.e. joint angles), Electromyography (EMG) data, and Ground Reaction Forces (GRFs) to infer predictions. 

In \cite{Sharma2022}, different feed-forward neural network configurations obtained through hyperparameter space exploration were trained and applied to estimate both joint biomechanical parameters (angles, reaction forces, torques) as well as muscle forces and activations in the upper extremities. The authors concluded that more complex neural architectures with optimal parameters obtained through hyperparameter space search, lead to more accurate estimations. Furthermore, RNN models pose as the most common approaches in human biomechanics estimation \cite{Loi2023, Mansour2023, Wang2023}. 
In particular, in \cite{Loi2023}, a Bidirectional Long Short-Term Memory network (BiLSTM), augmented with unsupervised domain adaptation layers, was proposed to perform domain alignment between experimental data from various movements, with the goal to generalize across multiple actions and align real-to-synthetic data simultaneously. 
LSTM has also been utilized for ankle, knee, and hip joint torque estimation in sit-to-stand trials, alongside other machine learning (ML) and deep learning (DL) techniques such as Linear Regression (LR), Support Vector Machines (SVM), and CNN, as documented in \cite{Mansour2023}, with LSTM indicating the best performance for the task. Similarly, LSTM and Gaussian Process Regression (GPR) were employed in \cite{Wang2023} to predict lower extremity joint torques during gait. 



\subsection{Physics-Informed Neural Networks}

Physics Informed Neural Networks, first introduced in \cite{Raissi2017, Raissi2019}, are standard machine/deep learning models that incorporate the underlying physical laws of their training dataset, expressed as partial differential equations (PDEs), in the learning process (i.e. loss function) to address both forward and inverse problems. Forward problems involve approximating the PDEs governing the training dataset to derive solutions, without prior knowledge of the ground-truth, relying solely on the provided input data and respective boundary conditions. State-of-the-art works in fluid \cite{Mahmoudabadbozchelou2022, Eivazi2024} and quantum mechanics/computing \cite{Sedykh2024, Trahan2024} utilize PINNs to solve forward problems, \comm{without explicitly modeling the underlying physics, unlike}instead of finite element \comm{and}or conventional physics-based methods. In inverse problems, PINNs leverage physics-based domain knowledge to penalize the estimation of physical quantities\comm{(i.e. limiting the solution space)}, enhancing robustness, accuracy, and generalization in cases of limited data availability as ground-truth, where conventional ML/DL methods are rendered ineffective. 

PINNs, despite being the revolution in the area of fluid and quantum mechanics, are also starting to be applied for biomechanics estimation tasks \cite{Zhang2023, Ma2024, ZhiboZhang2022, Taneja2024, Kumar2023, Zhang2022_2, Kumar2024, Zhang2023_2}. In \cite{Zhang2022_2, Zhang2023}, PINNs with CNNs as a base are used to predict muscle forces and joint angles based on EMGs, with the work in \cite{Zhang2022_2} developing a physics-based domain knowledge transfer technique to develop subject-specific PINN approaches. Furthermore, in \cite{Zhang2023_2} a distributed physics-informed DL approach is presented that builds local models acting on subdomains of the EMG input data to enhance the efficiency and robustness of muscle forces and joint angle estimation, while in \cite{Kumar2023} a PINN-based approach is explored for estimating the joint angle of the upper limb under various loads.
In the most recent works \cite{Taneja2024, Kumar2024}, involving the estimation of muscle parameters and joint angles \cite{Taneja2024} as well as joint torques \cite{Kumar2024} of the upper limbs, novel PINN architectures that incorporate Gated Recurrent Units (GRUs), are implemented. GRUs are a variant of RNNs with fewer training parameters and enhanced computational efficiency compared to LSTM models \cite{Cho2014, Chollet2017}, which significantly improve the accuracy of predictions involving time-dependent inputs and also support the learning of long-term time-dependencies. It is also worth mentioning the study in \cite{Ma2024}, which predicts muscle forces and identifies muscle-tendon parameters from unlabeled EMGs using both explicit and implicit PINN losses (i.e. describing the implicit relationship between muscle forces predicted by a DL model and those calculated by an embedded musculoskeletal model). 

\comm{
\subsection{Fatigue}
There have been numerous attempts to model fatigue in the Biomechanics literature. In [], fatigue is defined as a function of .... (insert function). 

Comment2: "fatigue is not usually considered in most of the existing MSK models, which leads to non-negligible errors" -> or data-driven motion synthesis frameworks

Moreover a model .., (3CC)
how to model fatigue using functions + 3CC (all relevant works from 2008 to 2020/1) + fatigue modeling using deep learning -> only \cite{Cheema2023} (Mocap of fatigued motion???)

Michaud2024 "Four-compartment muscle fatigue model to predict metabolic inhibition and long-lasting nonmetabolic components"}

\section{Methodology}
\label{sec:methodology}

\begin{figure*}[h!]
\centering
\includegraphics[width=\textwidth]{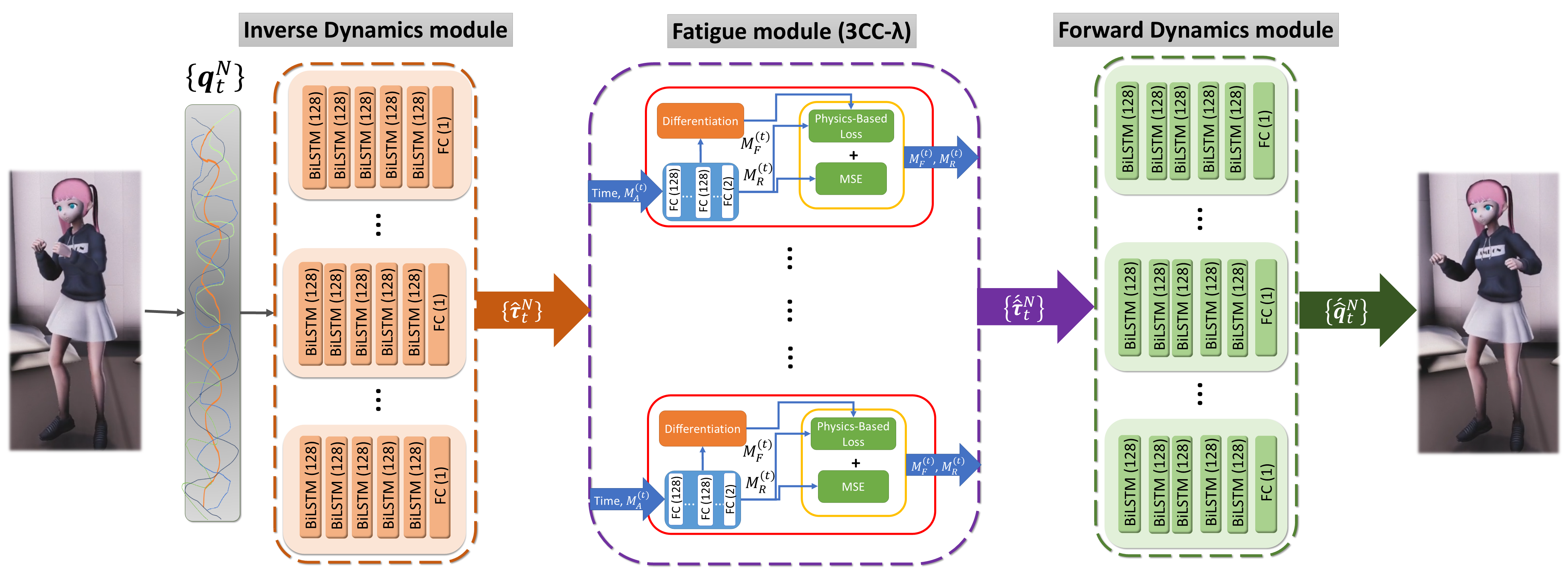}
\caption{General overview of our Fatigue-PINN framework.}\label{fig:1}
\end{figure*}

Fatigue-PINN is a PINN-based deep learning architecture to produce fatigued human motion. Our framework consists of three joint-specific modules: i) the ID module to account for the transition from the joint kinematics to the joint torque domain, ii) the Fatigue module, a PINN adaptation of the Three-Compartment-Controller, namely 3CC-$\lambda$, to simulate the effects of fatigue on the synthesized motion, and iii) an FD module to transform fatigued torques into fatigued joint angles, as depicted in Fig. \ref{fig:1}.  

In detail, in each motion frame $t$, the ID module is fed with a vector of $N$ joint angles $\mathbf{q}_t^N$, which in our case corresponds to 32 joint angles including both the joints of upper (shoulder, elbow, wrist) and lower limbs (hip, knee, ankle), pelvis, and lumbar and produces a vector of $N$  joint torques $\hat{\mathbf{\tau}}_t^N$. These torques are subsequently processed through the 3CC-$\lambda$ model (Fatigue module) to generate fatigued torques ($\acute{\hat{\mathbf{\tau}}}_t^N$), which are then given as input to the FD module to yield the fatigued motion (i.e. fatigued joint angles $\acute{\hat{\mathbf{q}}}_t^N$). 

Even though our ID / FD BiLSTM model accurately estimates multiple joint torques and angles simultaneously (e.g., achieving approximately $~86\%$ testing accuracy for upper body joint torques and angles as shown in Table \ref{tab4}), we opt for a joint-specific approach (as illustrated in Fig. \ref{fig:1}). This approach allows the neural network to focus all its learning capacity on a single task, resulting in enhanced accuracy, faster convergence, reduced training times, and lower computational requirements due to its reduced complexity.
More specifically, the ID/FD model extracts latent space correlations from all joint angles (torques) to effectively produce predictions solely for one joint. This method allows for handling motion artifacts on a joint level while enabling the setting of joint fatigue profiles as discussed in Section \ref{sec:3CC}. 

Supplementary to this Methodology section, is the accompanying Appendix, where more details concerning the procedure we followed to develop and train our Fatigue-PINN framework, are provided.

\subsection{Dataset \& Data Processing} 
\label{sec:dataset}

To train our BiLSTM ID and FD models, we utilized the \textit{Adapt Emotional Actions} open dataset, from the SIG Center for Computer Graphics
Multi-Modal Motion Capture Library of the University of Pennsylvania\footnote{Available online at: \url{https://fling.seas.upenn.edu/~mocap/cgi-bin/Database.php}}, that contains human motion capture data in C3D and BVH formats as well as force plate data (i.e. GRF data) from various activities, including knocking (a bell), box lifting, picking up a pillow, pointing to spot, punching, box pushing, throwing, walking and waving. 
The dataset is recorded over a subject's $11$ motion trials for each action, hence containing a total of $99$ motion trials, i.e. $9$ actions with $11$ motion trials each.

Each motion trial is expressed using a different emotion/behavior, based on the OCEAN model that provides a description of one's personality over five dimensions: Openness (imagination and creativity), Conscientiousness (organization, carefulness), Extroversion (social aspect of human character), Agreeableness (friendliness, generosity, etc.), and Neuroticism (emotional instability, tendency towards negative emotions), with each factor being bipolar ($+/-$) and composed of several traits as cited in \cite{Kapadia2013}. These personality traits, represented by the acronym OCEAN, are utilized to label each motion trial, meaning that the corresponding movement is performed to convey a specific emotion or behavior.
For example, in the motion trial "Waving\_E$+$\_02", the subject waves in a cheerful, energetic manner (i.e. E$+$ stands for Extroversion, thus, social, active, assertive, dominant, energetic behavior). Therefore, each movement consists of trials that are slightly different from one another, due to the expressed emotions, thereby introducing variability to the dataset within each motion. In particular, the standard deviation of the investigated joint angles for each different motion trial, as well as their standard deviation at every time frame across different emotions ranges from approximately 0.01 to 0.4. This variability is small, yet sufficient to enhance the generalization capability of our BiLSTM models across different variations of the same movement.      

Raw motion capture and GRF data are rotated to align with the OpenSim coordinate system and converted into TRC and MOT formats (motion files' formats) using Python scripts. These motion files are then utilized in the musculoskeletal modeling process in OpenSim to derive the ground-truth joint kinematics and corresponding joint torques for training our ID and FD models. The musculoskeletal model\footnote{Available online at: \url{https://simtk.org/projects/full_body/}} utilized in OpenSim is the one introduced in \cite{Rajagopal2016}, a Hill-type full-body model with muscle-actuated lower limbs (80 muscle-tendon units) and torque-actuated torso/upper limbs (17 ideal torque actuators) with 37 Degrees of Freedom (DoFs) to describe joint kinematics.   

The OpenSim analysis pipeline entails scaling the musculoskeletal model to match the anthropometric measurements of the subject used for capturing the Adapt Emotional Actions dataset. These measurements are obtained from the subject's neutral pose marker data. Subsequently, the inverse kinematics (IK) procedure is employed to calculate joint angles for both the upper (i.e. arm flexion/adduction/rotation, elbow flexion/pronation, and wrist flexion/deviation) and lower limbs (i.e. lumbar extension/bending/rotation, pelvis tilt/list/rotation, hip flexion/adduction/rotation, knee angle, ankle angle, and subtalar angle) based on the spatial trajectories of the markers. 
The computed joint angles, along with the experimentally measured GRF data are used to calculate joint torques via ID.

As for data pre-processing, the obtained joint kinematics and torques were normalized in the range $[0,1]$ using the min-max normalization method. Furthermore, it is worth noting that in the raw dataset, the duration (i.e. number of frames) of each motion trial varies, ranging from approximately $450$ to $600$ frames. Therefore, we resample each motion trial to the length of the longest trial using interpolation, ensuring that all emotional variations of an action have the same number of time steps. After interpolation, each motion trial has a duration of $\sim600$ frames, resulting in actions with motion trials with a total duration of $\sim6600$ frames.

\subsection{3CC}
\label{sec:3CC}



The Three-Compartment Controller, first introduced in \cite{Liu2002}, is a state machine that models the grouping of the muscle motor units of a human limb into three muscle activation states, as well as the transitions between them under static, \cite{Liu2002} or dynamic load conditions \cite{Xia2008, Frey-Law2012, Cheema2020, frey-law2021}. As depicted in Fig. \ref{fig:3CC}, at the beginning of a motion, motor units from the resting ($M_R$) state are transitioned to the active ($M_A$) compartment to satisfy the target load, $TL$ ("required force" to fulfill a specific task). If more motor units are activated than the ones required to meet $TL$, some motor units may return to the resting state (from the $M_A$ compartment to $M_R$). This bidirectional transfer of motor units between $M_A$ and $M_R$ is governed by a time-variant feedback controller, $C(t)$, which is dependent on TL as shown in Eq. (\ref{eq:Ct}). During activity, motor units inevitably end up in the fatigued state compartment, $MF$ (i.e. their force start to decay over time), at a fatiguing rate, $F$, and, then, they enter rest period, i.e. moved to $MR$, by a recovery rate of $R$. These transitions between the three states of 3CC in dynamic load conditions are described by the Equations (\ref{eq:MR}), (\ref{eq:MA}) and (\ref{eq:MF}) \cite{Xia2008, Frey-Law2012, frey-law2021}:

\begin{figure}[h!]
\centering
\includegraphics[width=\columnwidth]{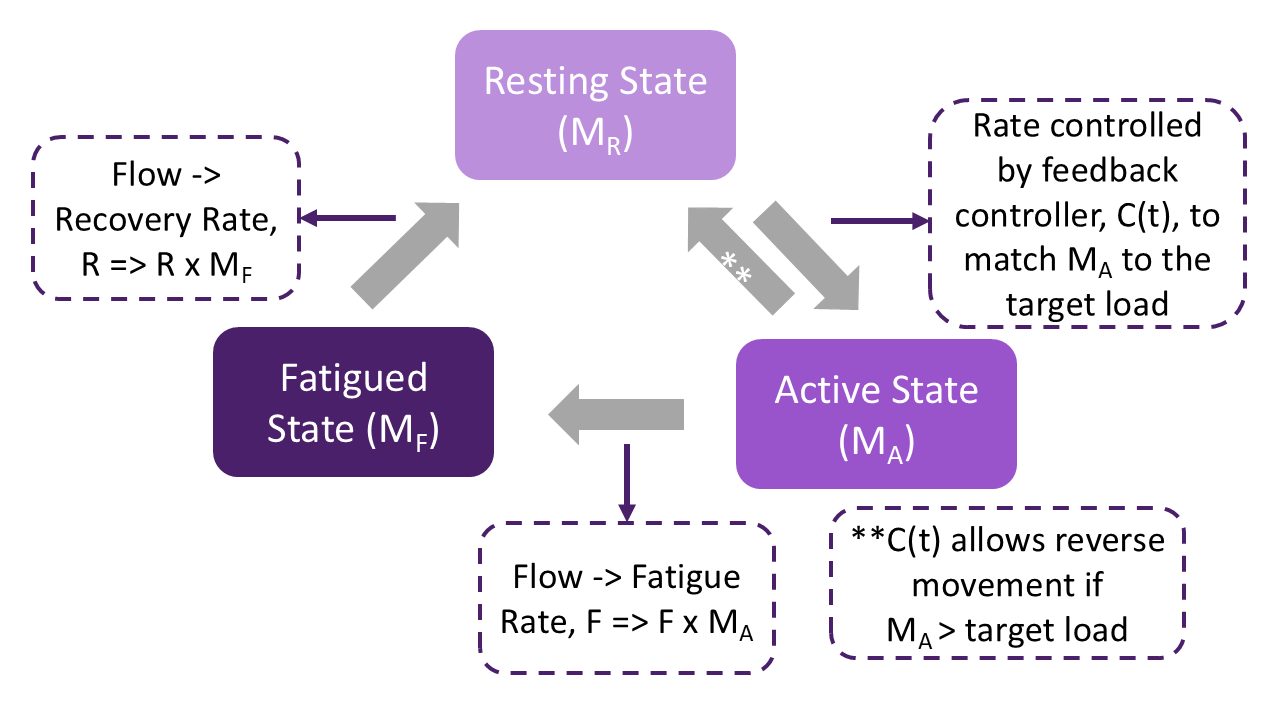}
\caption{3CC. Figure reproduced from \cite{Frey-Law2012}.}\label{fig:3CC}
\end{figure}

\begin{equation}
\frac{dM_R}{dt} = -C(t) + R * M_F
\label{eq:MR}
\end{equation}

\begin{equation}
\frac{dM_A}{dt} = C(t) - F * M_A
\label{eq:MA}
\end{equation}

\begin{equation}
\frac{dM_F}{dt} = F * M_A - R * M_F
\label{eq:MF}
\end{equation}

\begin{equation}
C(t) = 
\begin{cases}
     L_D * (TL - M_A) & {\scriptstyle  \text{if } M_A < TL \text{ and } M_R > (TL - M_A)}\\
    L_D * M_R & {\scriptstyle          \text{if } M_A < TL \text{ and } M_R \leq (TL - M_A)} \\
    L_R * (TL - M_A) & {\scriptstyle          \text{if } M_A \geq TL} \\
     \end{cases}
\label{eq:Ct}
\end{equation}

The values of $F$ and $R$ are specific for every joint, as can be found in the literature \cite{Frey-Law2012} and depend on the task the muscles perform. A greater ratio between parameters $F$ and $R$ accounts for faster accumulation of fatigue as reported in \cite{Xia2008, Frey-Law2012, frey-law2021}. Moreover, the parameters $L_D$ and $L_R$, which contribute to the formulation of $C(t)$ in Eq. (\ref{eq:Ct}), are constants that define, respectively, the ratio of muscle force development ($L_D$), thereby constraining the rate of muscle activation and the rate of muscle relaxation ($L_R$), ensuring the system's stability is preserved \cite{Xia2008, frey-law2021}.
 
Each one of the three compartments, $M_A$, $M_F$, and $M_R$ is expressed in $\%MVC$ – maximum voluntary contractions, and by summing the percentage of muscle motor units in each compartment, a total of $100\%$ is obtained. Residual Capacity ($RC$) quantifies the remaining muscle strength or maximum exerted force/joint torque capability as a percentage, accounting for the effects of fatigue, and, thus, is defined as follows \cite{Xia2008}:

\begin{align}
    M_A + M_F + M_R &= 100\% \label{eq:ALL}\\
    RC(t) = 100\% - M_F &= M_A + M_R \label{eq:RC}
\end{align}
    
Fatigue-PINN leverages a PINN version of 3CC, namely 3CC-$\lambda$, for modeling the fatigue state of specific muscle groups/joint actuator torques (w.r.t. the animated motion) during open human movements. As shown in Fig. \ref{fig:1}, our PINN 3CC-$\lambda$ is given as input the current active motor units $M_A$, which in our case are the joint torques of each DoF, $\hat{\mathbf{\tau}}_t^N$, produced by the ID model, and outputs the percentage of motor units in the complementary states, $M_F$ and $M_R$ in $MVC\%$. In contrast to previous work \cite{Cheema2023}, we compute the $RC$ of each joint, based on the $M_F$ state multiplied by a $\lambda$ factor (Eq. (\ref{eq:RCC_hat})) that empirically describes the fatigue profile of each muscle/joint to better align joint fatigue in different movements and allow for joint-specific fatigue configurations. The $\lambda$ factor, ranging between 0 and 1, is strategically selected to mitigate the influence of fatigue on motion by a small percentage, hence producing smoother motions. This parameter is action-dependent and varies for each joint. Based on Eq. (\ref{eq:3CC-lambda}), joints whose fatigue profile is set to  $\lambda = 0$ will not contribute to the fatigued motion (fatigued torques will not be applied upon them), while a joint with $\lambda = 1$, has its $RC$ reduced by the $M_F$ factor. In the latter case, the fatigued torques applied to the respective joints are completely dependent on the 3CC, i.e. the ones arising from the fatigue state. Subsequently, we apply $RC$ as a time-varying factor to simulate the gradual decaying of the maximum exerted joint torques to directly generate the fatigued ones, $\acute{\hat{\mathbf{\tau}}}_t^N$, as in \cite{Xia2008, Cheema2023} (Eq. (\ref{eq:3CC-lambda})):

\begin{align} 
    \hat{RC(t)} &= 100\% - \lambda * M_F \label{eq:RCC_hat}\\ 
    \acute{\hat{\mathbf{\tau}}}_t^N &= \hat{RC} * \hat{\mathbf{\tau}}_t^N \label{eq:3CC-lambda}
\end{align} 

For instance, if $M_F = 20\%$, at elbow joint at frame t, and $\lambda = 0.6$ then from Eq. (\ref{eq:RCC_hat}) we derive that the maximum exerted elbow torques peak at $RC(t) = 88\%$ (instead of $RC(t) = 80\%$), meaning that $\lambda$ factor, contributes in smoothly injecting joint-specific fatigue and producing more physically-plausible motions without intermittent ("step-like") movements.


As depicted in Fig. \ref{fig:1}, the 3CC-$\lambda$ model consists of five fully connected layers, being fed with the current time frame $t$ and current torque of each DoF, $\hat{\mathbf{\tau}}_t^N$ (i.e. $M_A$), and produces $M_F$ and $M_R$ of the same frame.  
Its loss, $L$ as formulated in Equations (\ref{eq:3CC-loss1}), (\ref{eq:3CC-loss2}), and (\ref{eq:3CC-loss3}), is comprised of two terms (in accordance with Fig. \ref{fig:1}) i) the Mean Squared Error (MSE) loss of the neural network, $L_{NN}$, and ii) the physics-based loss, $L_{PB}$, based on Equations (\ref{eq:MF}) and (\ref{eq:MR}) that describe the transition between the active and fatigued state and the transition between the fatigued and the resting state, respectively: 

\begin{equation}
     L = L_{NN} + L_{PB} 
     \label{eq:3CC-loss1}
\end{equation} 
\begin{equation}
    L_{NN} = \frac{1}{T}\sum_{t=0}^{T}(M_F - \hat{M_F})^2 + \frac{1}{T}\sum_{t=0}^{T}(M_R - \hat{M_R})^2
    \label{eq:3CC-loss2}
\end{equation}
\begin{align}
     L_{PB} = \frac{1}{T} \sum_{t=0}^{T}(\frac{d\hat{M_F}}{dt} - F * M_A + R * \hat{M_F})^2 + \notag\\
    \frac{1}{T} \sum_{t=0}^{T}(\frac{d\hat{M_R}}{dt} + C(t) - R * \hat{M_F})^2
    \label{eq:3CC-loss3}
\end{align}

where $\hat{M_F}$ and $\hat{M_R}$ represent the percentage of motor units at the fatigued and resting states as estimated by the neural network and $M_F$ and $M_R$ correspond to the ground-truth values. 

To simulate the flow between the compartments of 3CC, we train our PINN using datasets referenced in \cite{frey-law2021}, such as the one in \cite{Burnley2009}. Such datasets, explore the fluctuations of joint actuator torques after a fatiguing task based on experimental setups (e.g. measuring joint torques after a series of MVCs of a corresponding muscle set). Given the very few data in these datasets (i.e. $\sim40-50$ time frames), the physics-based loss penalizes the neural network's predictions, thus, limiting the solution space and better guiding the learning process. In simpler words, these datasets aid the PINN model to learn the functionality of the 3CC faster, leading to accelerated convergence (typically within $\sim50$ epochs). As mentioned in Section \ref{seq:Ablations}, the 3CC-$\lambda$ network can be also trained unsupervisingly, where only the loss term $L_{PB}$ contributes to the training procedure.

Furthermore, $F$ and $R$, are set to values specific for every joint upon which the 3CC-$\lambda$ is applied. For example, for the elbow joint, the parameters are set to $F = 0.00912$ and $R = 0.00094$ values, according to \cite{Frey-Law2012}. Constants $L_D$ (rate of muscle force development) and $L_R$ (rate of muscle relaxation) are set to $10$ for all joint configurations \cite{Xia2008}. It is worth mentioning that $L_D$ and $L_R$ have the least influence on the predicted fatigue since the temporal progression of muscle force development and relaxation factor is negligible to the one of fatigue as explained in \cite{Xia2008}. The training of 3CC-$\lambda$ was conducted using Python's Keras library \cite{chollet2015keras} employing the adaptive moment estimation (Adam) \cite{Kingma2014} optimizer with a batch size of 32 and a learning rate of $0.001$, to improve the training convergence and prevent gradient descent from getting stuck at local minima.

\subsection{Inverse/Forward Dynamics Model}

In both motion synthesis and musculoskeletal dynamics estimation research, RNNs are preferred over other machine learning methods. This preference arises from their ability to extract temporal features from motion sequences, which is essential for motion and dynamics estimation (within the same time frame $t$), prediction (for the subsequent time frame $t+1$), and the synthesis of novel movements. 

In this work, we exploit Bidirectional Long Short-Term Memory neural networks to develop both our ID and FD surrogate models, which produce precise predictions without requiring GRFs for their training. BiLSTMs possess two parallel sequences of forward and backward feedback connections, in contrast to LSTMs, which have only a single forward loop. This dual-sequence architecture enables BiLSTMs to handle both past and future time dependencies effectively, preventing gradients from gradually vanishing during training and thereby producing more accurate predictions \cite{Schuster1997, Ihianle2020}. 

\begin{figure}[h!]
\centering
\includegraphics[width=\columnwidth]{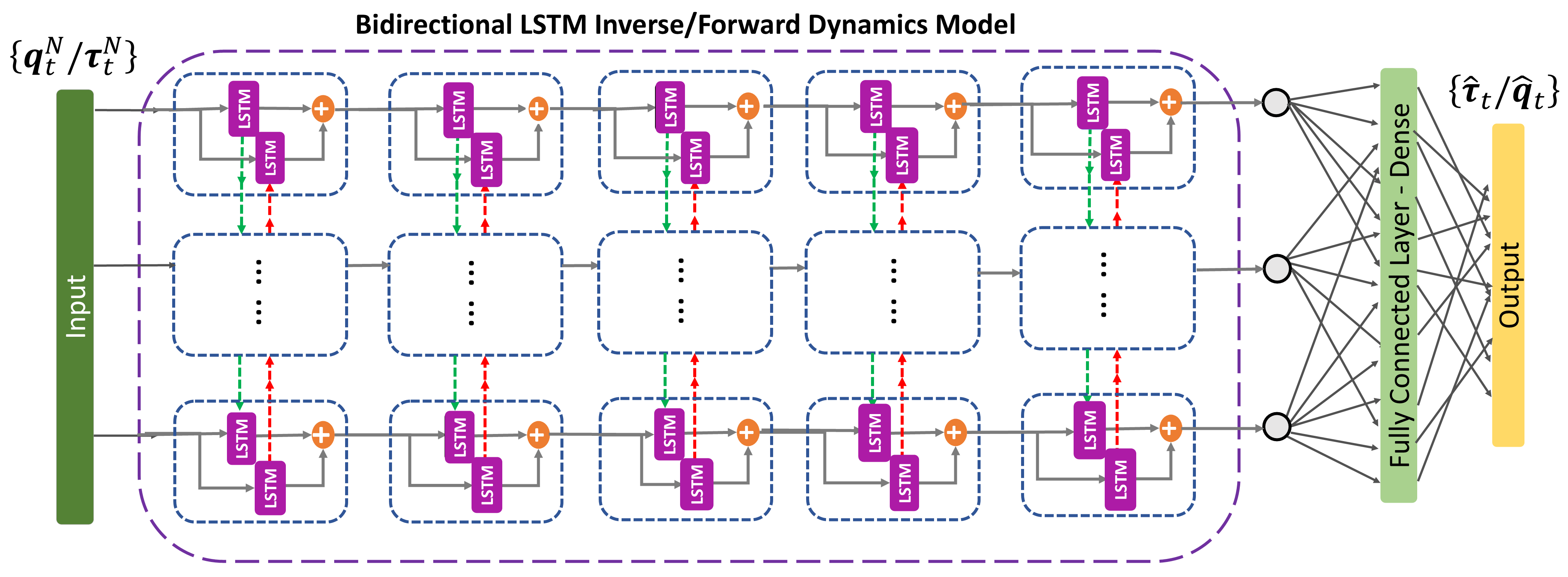}
\caption{BiLSTM Inverse/Forward Dynamics Model. The feedback loop of each BiLSTM layer is marked with green arrows, whereas its backward sequence is indicated with red arrows.}\label{fig:2}
\end{figure}

The forward and backward sequence outputs are computed using Equations (\ref{eq:1}) and (\ref{eq:2}), which are the standard LSTM updating equations \cite{Schuster1997, Cui2018, Ihianle2020}. In particular, the output $\overrightarrow{h}_t$ of the forward sequence is computed iteratively in a positive time direction spanning from $t = T$ to time $t = 1$. On the contrary, the backward loop is "looking" at the past, thus, its output $\overleftarrow{h}_t$ is calculated in a negative time direction from $t = 1$ to $t = T$ (see Fig. \ref{fig:2}):

\begin{equation}
    \overrightarrow{h}_t = g (U_{\overrightarrow{h}}x_t + W_{\overrightarrow{h}}\overrightarrow{h}_{t-1} + b_{\overrightarrow{h}})
\label{eq:1}
\end{equation}

\begin{equation}
    \overleftarrow{h}_t = g (U_{\overleftarrow{h}}x_t + W_{\overleftarrow{h}}\overleftarrow{h}_{t-1} + b_{\overleftarrow{h}})
\label{eq:2}
\end{equation}

\noindent where $x_t$, is the input data $x$ at time frame $t$, $U$ and $W$ are the weight matrices and $b$ is the bias vector. The input of the ID model is defined as $x_t = \mathbf{q}_t^N$, i.e. a vector of $N$ joint angles at frame $t$, while $x_t = \hat{\mathbf{\tau}}_t^N$ is the input for the FD model, i.e. a vector of $N$ fatigued torques at time $t$. The values of the weight matrices and the bias vector are updated ("learned") during training. 


Furthermore, the BiLSTM layer produces an output vector at each time frame $t$, $y_t$, which is computed using the calculated feedback sequences' outputs, $\overrightarrow{h}_t$ and $\overleftarrow{h}_t$, as indicated by Eq. (\ref{eq:3}) \cite{Ihianle2020}. 

\begin{equation}
    y_t = g (V_{\overrightarrow{h}}\overrightarrow{h}_t + V_{\overleftarrow{h}}\overleftarrow{h}_t + b_y)
\label{eq:3}
\end{equation}

\noindent where in our case $g$ is a concatenating function used to combine the outputs of the two feedback sequences \cite{Cui2018}. Given that $y_t$ is the output of the last BiLSTM layer in each of the ID/FD architecture, it is then passed to a fully connected layer (as illustrated in Figures \ref{fig:1} and \ref{fig:2}). In our joint-specific framework, the fully connected layer in the ID model computes the current torque for one joint denoted as $\hat{\mathbf{\tau}}_t$, while the fully connected layer of the FD architecture produces a single fatigued joint angle, $\acute{\hat{\mathbf{q}}}_t$, at frame $t$. 

As shown in Figures \ref{fig:1} and \ref{fig:2}, the ID and FD models utilize a symmetric BiLSTM architecture consisting of five BiLSTM layers each one of them containing 128 units with linear activation on the first BiLSTM layer and ReLU activations on the rest hidden layers. Additionally, a Fully Connected layer with linear activation is included as the output of the ID/FD models. Similarly to the training of 3CC-$\lambda$, both ID and FD BiLSTM networks were trained in Python Keras, utilizing the Adam optimizer with a learning rate of $0.001$. The learning process was conducted with a batch size of 32 and an MSE loss function, over 1000 epochs, with early stopping enabled to prevent the model from over-fitting. Our joint-specific approach, meaning training a Fatigue-PINN model separately for each joint angle, may entail over-fitting. Producing predictions (either torques or angles) only for one joint requires extracting latent-space dependencies for one output, which results in faster convergence, i.e. in fewer epochs, rendering early stopping essential. Early stopping also tackles over-fitting issues due to the lack of motion capture data from multiple subjects.

The final models and hyperparameters were selected after assessing various architectures for our ID/FD BiLSTM models in terms of the number of BiLSTM layers and units, batch size, and learning rate. In Table \ref{tab11}, the MSE values of these configurations are reported indicatively for the punching motion type, to assess the testing accuracy of every architecture. MSE measures the average squared difference between the ground-truth values and the predicted values. During the computation of this metric, the errors are squared, meaning that larger errors can significantly increase the MSE and indicate a deterioration in performance.

\begin{table}[ht]
\setlength\tabcolsep{1.5pt}
\centering
\begin{threeparttable}
\caption{Comparison Across ID/FD BiLSTM Architectures}
{\begin{tabular}[t]{lcccc}
\toprule
\multicolumn{5}{c}{Variations in Number of BiLSTM Layers (128 units, 32 batch size, lr = $0.001$)} \\
\midrule
Model & 3 layers & 4 layers & 5 layers & 6 layers \\
\midrule
ID (MSE) & $0.043$ & $0.060$ & $\textbf{0.028}$ & $0.11$  \\
FD (MSE) & $0.88$ & $0.20$ & $\textbf{0.152}$ & $0.27$  \\
\midrule
\multicolumn{5}{c}{Variations in Number of Units (5 layers, 32 batch size, lr = $0.001$)}\\
\midrule
Model & 64 units & 128 units & 256 units & 512 unit\\
\midrule
ID (MSE) & $0.048$ & $\textbf{0.028}$ & $0.053$ & $0.036$\\
FD (MSE) & $0.91$ & $\textbf{0.152}$ & $0.21$ & $0.26$\\
\midrule
\multicolumn{5}{c}{Variations in Batch Size (5 layers, 128 units, lr = $0.001$)}\\
\midrule
Model & 16 batch size & 32 batch size & 64 batch size & 128 batch size\\
\midrule
ID (MSE) & $0.047$ & $\textbf{0.028}$ & $0.060$ & $0.035$\\
FD (MSE) & $0.65$ & $\textbf{0.152}$ & $0.87$ & $0.23$\\
\midrule
\multicolumn{5}{c}{Variations in Learning Rate (5 layers, 128 units, 32 batch size)}\\
\midrule
Model & lr = $0.01$  & lr = $0.001$ & lr = $0.0001$ & lr = $0.00001$\\
\midrule
ID (MSE) & $0.86$ & $\textbf{0.028}$ & $0.032$ & $0.050$\\
FD (MSE) & $7.3$ & $\textbf{0.152}$ & $0.18$ & $0.21$\\
\bottomrule
\end{tabular}}
\label{tab11}
\begin{tablenotes}
\item[$^a$] {A comparison between different ID/FD BiLSTM architecture and hyperparameter combinations in terms of Mean Squared Error (MSE). Evaluations on variations of a hyperparameter were conducted independently of the rest of the hyperparameters of an architecture. For example, when imposing changes in the number of BiLSTM layers on a model, the rest of the hyperparameters remain unchanged (e.g., 128 units, 32 batch size, lr = 0.001). All architectures were evaluated indicatively on the punching motion type. The architectures indicating the best performance are marked in bold. The abbreviation "lr" stands for learning rate. All values are multiplied by $10^{-2}$.}
\end{tablenotes}
\end{threeparttable}
\end{table}%

From Table \ref{tab11} it is clear that the architecture chosen for our ID/FD models, namely 5 BiLSTM layers with 128 units each, batch size = $32$, and learning rate = $0.001$, has the best MSE values compared to the rest of the models that we evaluated. It is worth mentioning that these MSE values are measured over the normalized data, ranging from 0 to 1. Therefore, an MSE of $0.00152$ resulting in an RMSE = $\sqrt{0.00152} \sim= 0.0389$, shows that the predictions of our model differ on average by $0.0389$ from the ground-truth values. Thus, smaller MSE values indicate a minimal error and, as a result, a high accuracy. Overall, a small enough batch size enables the model to learn over small parts of the training dataset at each iteration, while a learning rate of $0.001$, which is commonly used in literature, offers a good frequency of updating the model's weights without significantly increasing training time (smaller learning rate values result in updating weights more often, hence larger training times). Moreover, utilizing 128 units provides a good balance between accuracy and the number of training parameters as well as training time (i.e. the more units used for each layer, the more the parameters). Similar MSE values were yielded by testing the same architectures on throwing and waving motion types.


\section{Results}

\begin{figure*}[h!]
\centering
\includegraphics[width=\textwidth]{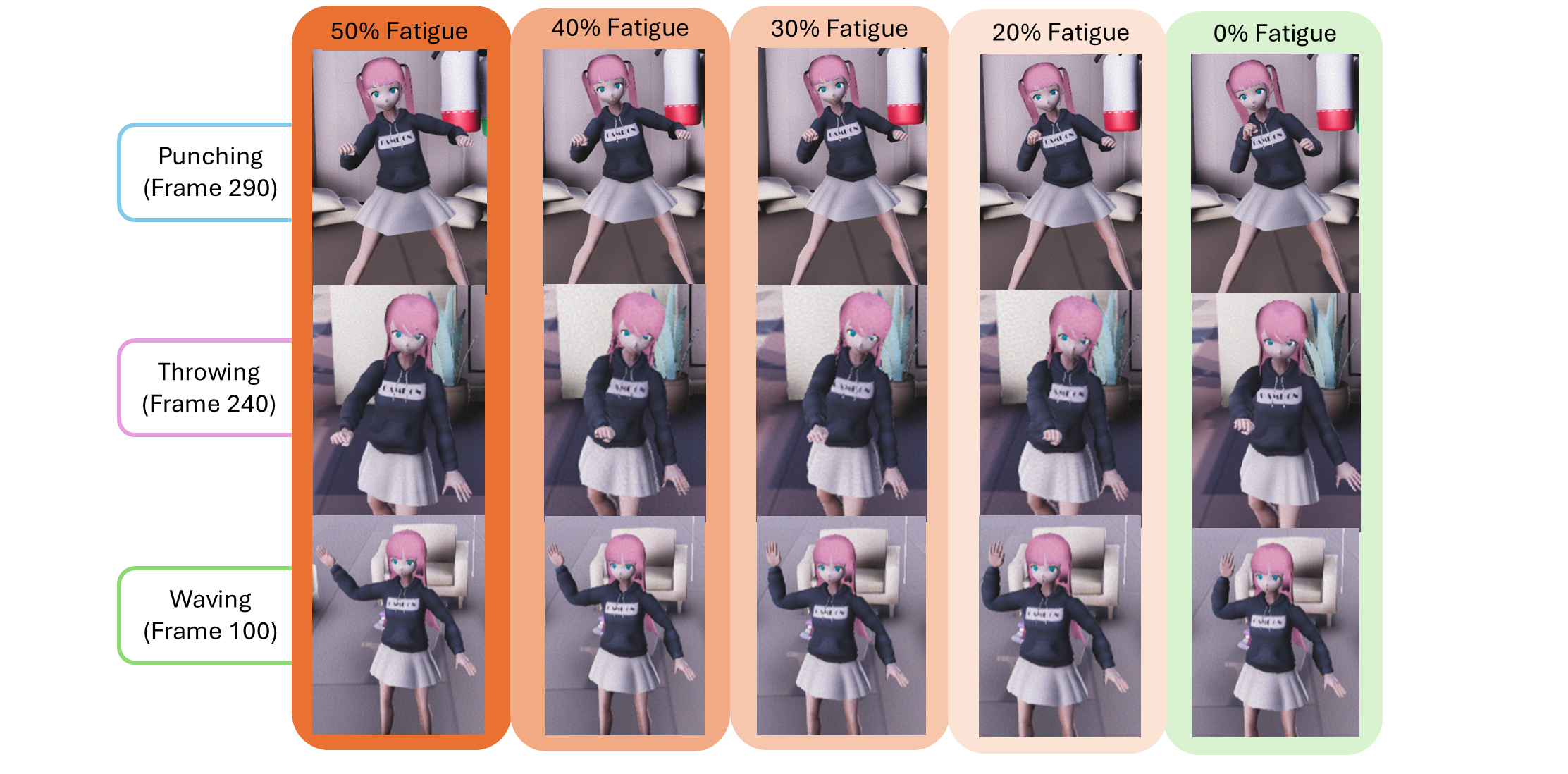}
\caption{The impact of different levels of fatigue on punching (Frame 290, i.e. the moment of hit), throwing (Frame 240, at throw), and waving (Frame 100) motions. The levels of fatigue are defined w.r.t Residual Capacity as arises from Eq. (\ref{eq:3CC-lambda}). For instance, $30\%$ fatigue $\rightarrow RC(t) = 70\%$, etc.} 
\label{fig:9}
\end{figure*}

\comm{
\textcolor{red}{TO DO: \begin{itemize}
    \item Video me oles tis kinhseis kai isws diorthwsh artifacts
    \item "Real-time" dinw mia dikh moy kinhsh kai paragei me to slider mia diaforetikh fatigued kinhsh
    \item citations apo TVCG
    \item diagrafh references (kamia 20aria)
\end{itemize}}}

\subsection{Fatigue Effects in Animation}
\label{sec:fatigue_effects}

The proposed Fatigue-PINN framework was trained and tested upon open-type movements (as defined in this work), i.e. waving, punching, throwing, to assess its capability of effectively producing fatigued motion in various action classes (see Fig. \ref{fig:9}). It is worth mentioning that we train a Fatigue-PINN framework for each one of the aforementioned motion types. Therefore, for each action, we performed cross-trial validation to enhance the robustness and accuracy of the model across different motion trials (i.e. emotional variations of the same movement).

We opted to apply fatigued torques on the upper body (shoulders, elbows, wrists, and lumbar), whose motion is crucial for perceiving the effects of fatigue in these particular open motions. Therefore, the fatigue profiles -$\lambda$ values- for the upper body joints are set between $0.6$ to $1$ across different motion types (punching, waving, and throwing), to slightly reduce the influence of fatigue (i.e. produce smoother movements), yet retain the effects of the fatigue state produced by the 3CC on the resulting motion. The $\lambda$ factor values for the lower body joints in all scenarios presented in this work are set to 0.

Our results are simulated in Unity Game Engine Environment \cite{Unity}, utilizing the \textit{Anime Girl Character}\footnote{Available online at: \url{https://assetstore.unity.com/packages/3d/characters/humanoids/casual-1-anime-girl-characters-185076#publisher}} avatar from Unity Asset Store. 


\begin{figure}[h!]
\centering
\includegraphics[width=\columnwidth]{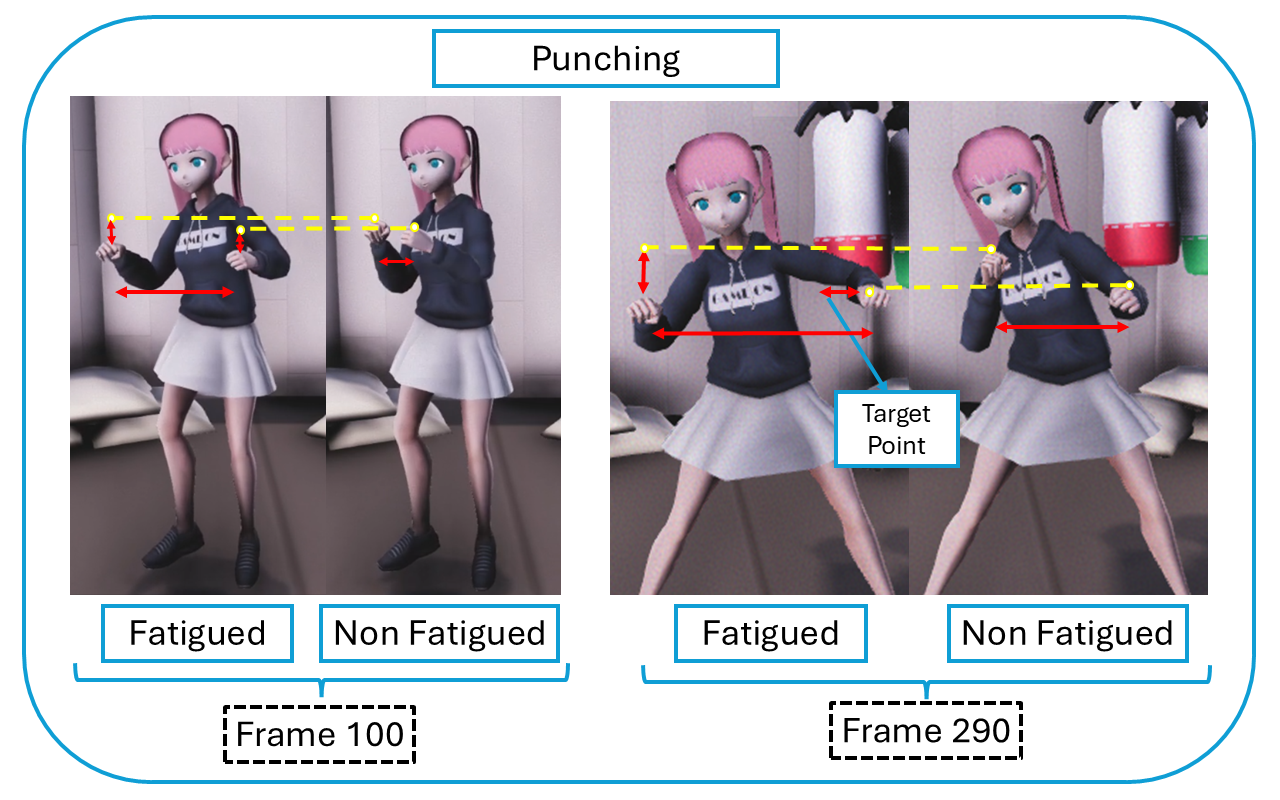}
\caption{Effects of fatigue on punching motion. At frame $100$ the character is at guard stance, and frame $290$ exhibits the moment of impact of the left hand. The discrepancies in hand positioning are marked with yellow dotted lines, while the distances between fatigued and non-fatigued hand positions are marked with red arrows.}\label{fig:4}
\end{figure}

\begin{figure*}[h!]
\centering
\includegraphics[width=\textwidth]{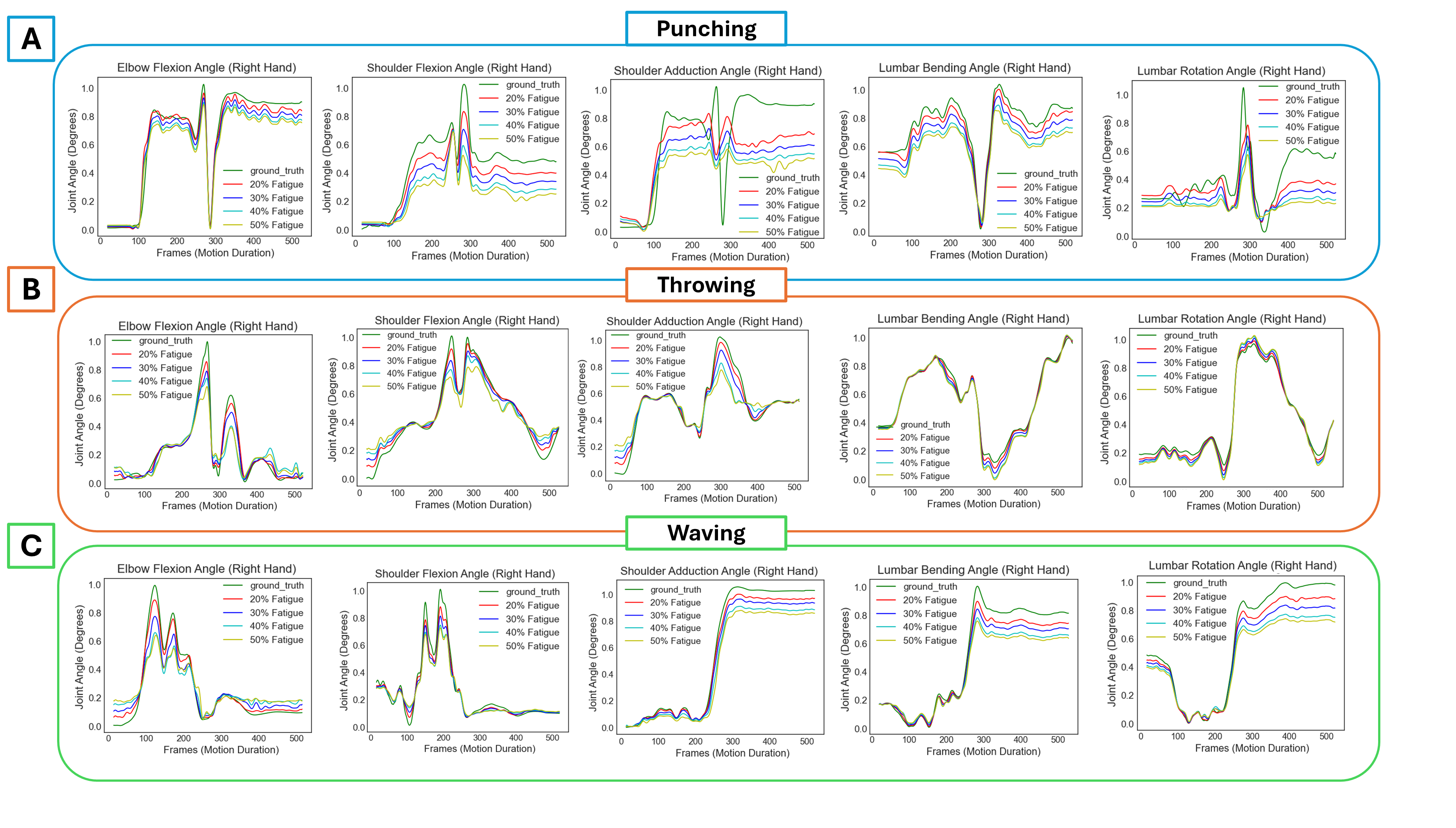}
\caption{The impact of fatigue on lumbar bending and rotation, as well as elbow flexion, shoulder flexion, and shoulder adduction angles of the right arm during the movements of punching, throwing, and waving (Sections A, B, and C, respectively). The effects of various levels of fatigue are demonstrated; for instance, $20\%$ of fatigue corresponds to the exertion of $80\%$ of maximum torque from the respective joints. The results are outlined normalized in the range of $[0,1]$.}\label{fig:5}
\end{figure*}

In both motion frames depicted in Fig. \ref{fig:4}, the character on the left exhibits fatigued movement ($RC(t)=\sim55\%$) in all joints of both arms, while the right is the non-fatigued one. In punching motion (Fig. \ref{fig:4}), the hand positioning is different than in the non-fatigued motion, rendering the fatigued character "unguarded" to hits (see frame $100$ in Fig. \ref{fig:4}). Moreover, in frame $290$ as shown in the right section of Fig. \ref{fig:4}, another discrepancy in motion due to fatigue is apparent, with both arms performing a wider movement while hitting, thus, missing the target. 
These two effects on the kinesiology of the 3D character, guard drop, and non-accurate punches, are behaviors observed alongside the increase of perceived fatigue in boxing athletes \cite{Dunn2017, Dunn2019}, while guard drops pose as a tactical strategy to mitigate fatigue by providing a small period of rest \cite{Dunn2017}. The latter, known as a pacing strategy, involves the adaptation of exercise intensity (decrease/increase effort accordingly) to minimize the impact of fatigue and thereby attain the desired outcome \cite{Smits2014} (e.g. determine the winner in a boxing bout). Moreover, the "unguarded" stance is a result of increased shoulder abduction, therefore, decreased shoulder adduction as seen in Section A of Fig. \ref{fig:5}, and increased elbow extension (i.e. less flexed elbow as in Fig. \ref{fig:5}). Notably, while shoulder adduction generally decreases, it increases at the moment of hitting impact (around frame $250$), resulting in a corresponding decrease in shoulder abduction. These findings align with those in \cite{Haralabidis2020}. Lumbar bending and rotation are reduced, with the character inclining to the left during striking with the right hand, while the upper body exhibits greater rotation to the right. These discrepancies in joint angles are reflected in the respective torques, which are expected to be reduced in the presence of fatigue by the 3CC-$\lambda$ model. 

\begin{figure}[h!]
\centering
\includegraphics[width=\columnwidth]{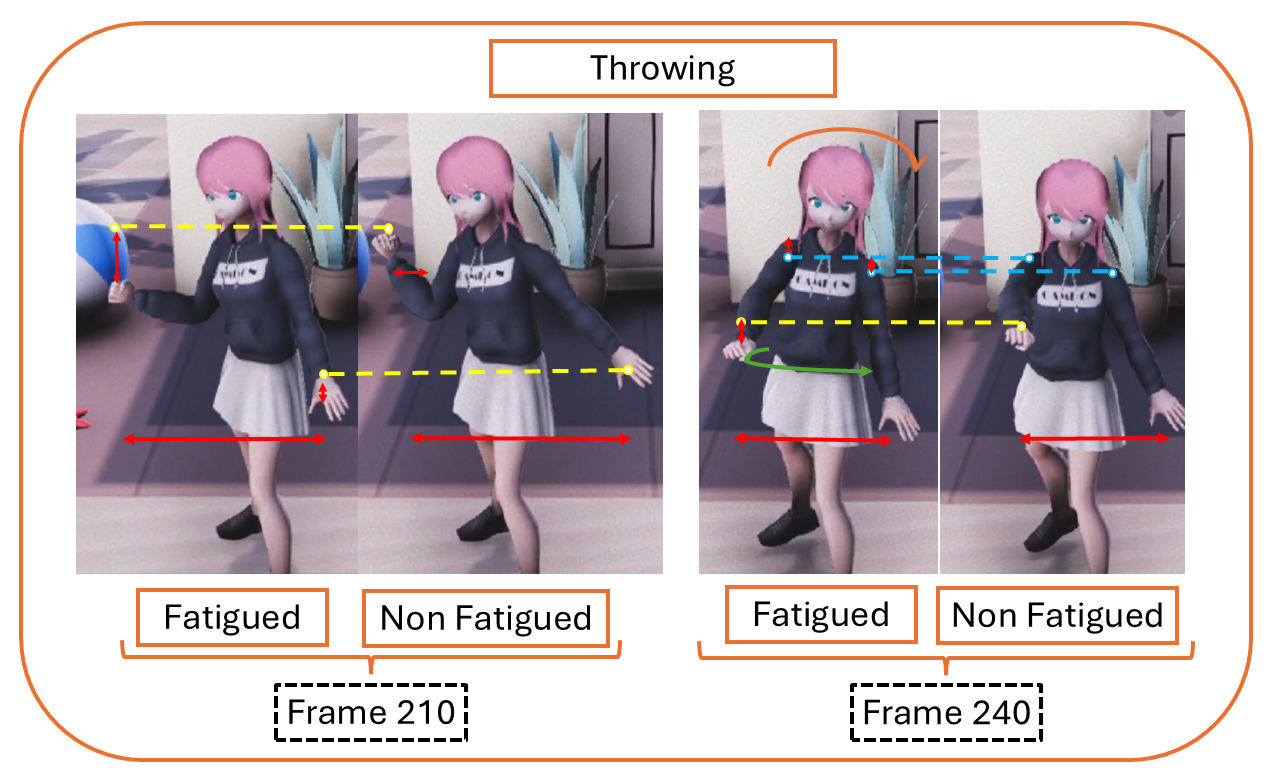}
\caption{Effects of fatigue on throwing motion. Frame $210$ indicates the pre-throwing phase, and frame $240$ depicts the moment of throwing. The changes in hand positioning between fatigued and non-fatigued motion are marked with yellow dotted lines, while the discrepancies in shoulder positions are marked in cyan dotted lines. Moreover, the increased flexion towards the left side (i.e. decreased lumbar bending angle in the lateral plane) is noted with an orange arrow, whereas a green arrow illustrates lumbar left rotation.}\label{fig:6}
\end{figure}

\begin{figure}[h!]
\centering
\includegraphics[width=\columnwidth]{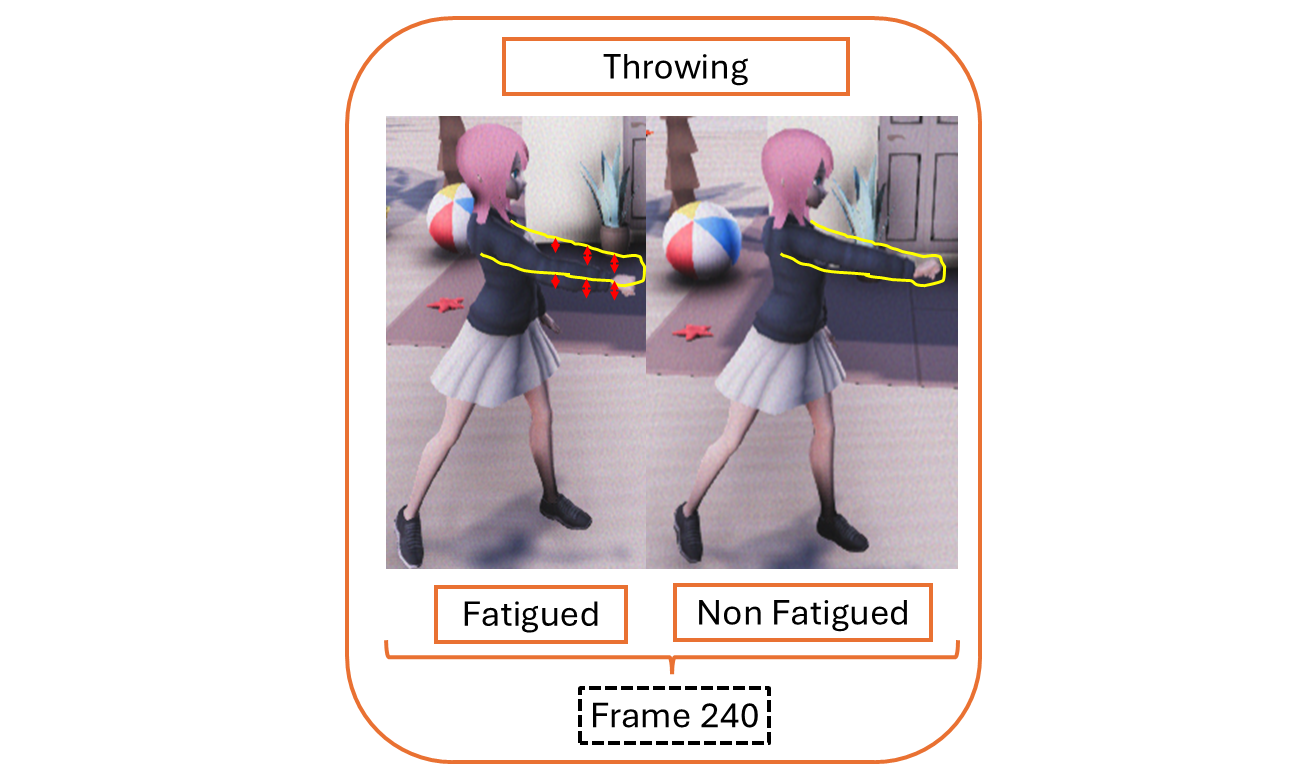}
\caption{Profile shots of throwing motion at frame $240$ (moment of throwing). At throw, the right shoulder and elbow are positioned more posterior, while the arm is positioned lower due to decreased shoulder flexion.}\label{fig:7}
\end{figure}

Similar results are observed in throwing (see Fig. \ref{fig:6}), with the right hand, the one performing the motion, being directed in a lower position than the one intended. The externally perceived impact of fatigue in this movement can be associated with the one in pointing tasks \cite{Fuller2009, Yang2019}, where one hand extends towards a specific position while the other remains relatively inactive. However, throwing is inherently more physically demanding, thereby we intuitively expect it to lead to a more pronounced fatigue effect. Regarding the changes in position as depicted in the right section of Fig. \ref{fig:6} (frame $240$), the shoulder joint positions of both arms are moving towards the non-throwing side, while the right shoulder is positioned higher at throw, which is in accordance with the findings in \cite{Fuller2009}, and may provide compensation mechanisms for the increased shoulder elevation and elbow extension \cite{Fuller2009, Grantham2014}. Moreover, following fatigue, the shoulder and elbow joints of the right arm (the one performing the throwing/pointing task) adopt a more posterior position, indicating a posture that reduces shoulder torques \cite{Fuller2009} (Fig. \ref{fig:7}), consistent with the predictions of the 3CC-$\lambda$ model. 

Similar to the findings in \cite{Fuller2009, Yang2019}, the fatigued character in the left of Fig. \ref{fig:7} exhibits reduced shoulder flexion of the right hand at the moment of throwing (frame $240$). Akin to the results in \cite{Fuller2009}, upper-body inclination towards the non-reaching arm's side due to decreased lumbar bending angle (i.e. more flexion in the lateral plane towards the left side) is also evident at throw (Frame $240$ in Fig. \ref{fig:6}). Moreover, in accordance with \cite{Yang2019}, greater lumbar rotation (i.e. increased lumbar left rotation angle) throughout throwing (frames $210 - 300$) is apparent in fatigued movement as illustrated in Figures \ref{fig:6} and \ref{fig:5} (Section B). The above kinematic effects are attributed to elbow and/or shoulder fatigue, as observed in \cite{Fuller2009, Cowley2017, Yang2019}.


Furthermore, the right-hand elbow is less flexed (reduced elbow flexion angle as seen in Figures \ref{fig:6} and \ref{fig:5}), which is considered a mean of compensation for both elbow and trunk fatigue as mentioned in \cite{Yang2019}, whereas elbow fatigue also influences shoulder abduction/adduction, leading to increased abduction (i.e. decreased adduction as shown in Fig. \ref{fig:5}) \cite{Yang2019}. 

\begin{figure}[h!]
\centering
\includegraphics[width=\columnwidth]{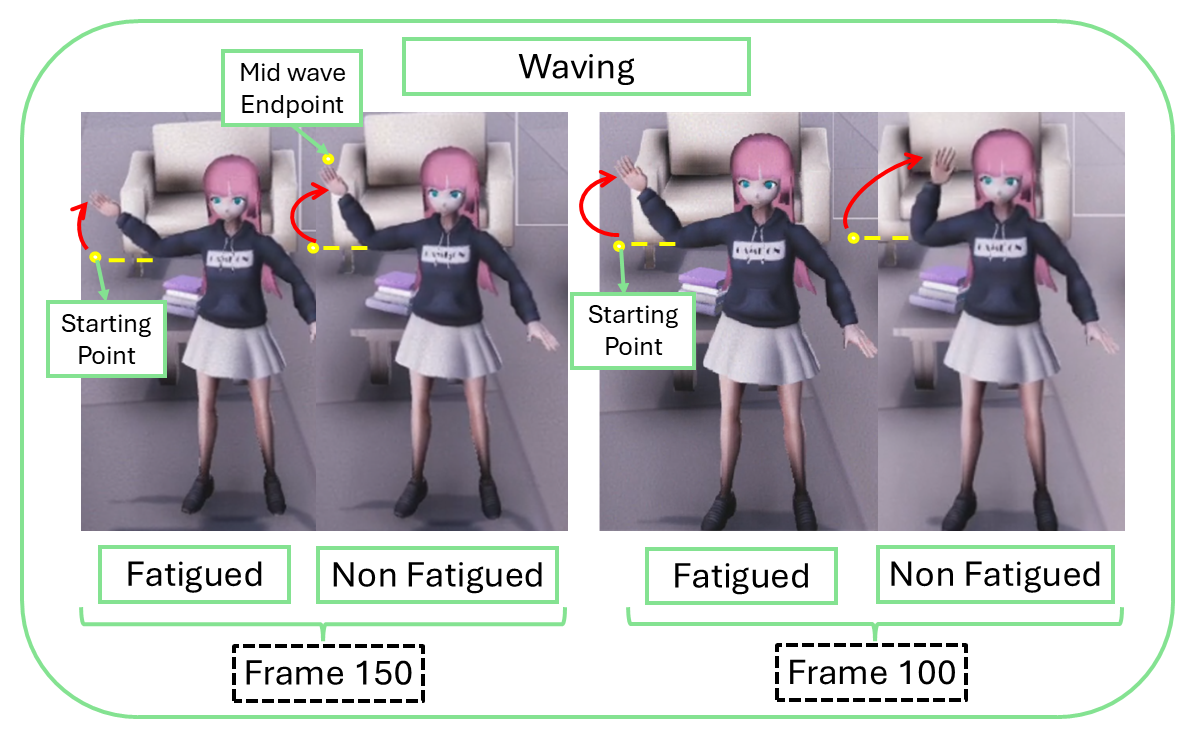}
\caption{Effects of fatigue at waving motion. The starting and mid wave points of the movement are indicated by yellow markers, while red arrows denote the angle between the waving right hand and the starting point. }\label{fig:8}
\end{figure}

The effects of fatigue are also evident in waving movement, where the elbow of the right (i.e. waving hand) is rendered more extended, resulting in a limited range of motion (Section C in Fig. \ref{fig:5} and Fig. \ref{fig:8}). Overall, the greater the decrease in the percent of maximum exerted torques (i.e. greater $RC(t)$ as arises from Eq. (\ref{eq:3CC-lambda})) the more the influence of fatigue in the produced motions.\footnote{Please refer to the video in the supplemental material for more qualitative results regarding our framework.} 

\subsection{Perceptual Study}

To further evaluate the fatigued motion sequences produced by Fatigue-PINN in terms of conveying externally perceived fatigue, we conducted a perceptual study. The study included a questionnaire drafted based on the existing literature \mbox{\cite{Darici2023, Dehesa2020, Edstrom2020}} that describes good practices for performing visual quality evaluations on animation videos, and especially on the work of the Disney Research Team \mbox{\cite{Vicovaro2014}}, where edited throwing animations were assessed in terms of perceptual plausibility. In accordance with the above works, we constructed a questionnaire in which participants were exposed to short videos (of duration $\sim30-40$ seconds), showing fatigued (depicting various levels of fatigue) and non-fatigued movements, and were asked to:
\begin{itemize}
    \item identify which of these videos they perceive as the most fatigued,
    \item specify which fatigued effects they consider are conveyed through the fatigued animation videos,
    \item assess the fatigued motion sequences in terms of realism/physical-plausibility, as well as
    \item provide their feedback regarding what they expect to see from a fatigued motion of X action (i.e. punching, throwing, or waving). 
\end{itemize}

Our study group consists of $19$ participants, among which $13$ ($68.4\%$) are male and $6$ ($31.6\%$) are female, while a "Prefer not to say" option was included to preserve the right of the participant not to identify with a binary gender. The age group of the participants is $18-44$, with more than half ($11$ - $57.9\%$) of the participants belonging in the range of $25-34$ years old. We also asked the participants to specify their experience in research areas relevant to this work, namely Animation, Computer Graphics, and Biomechanics, to obtain feedback from both inexperienced and individuals who are knowledgeable in motion synthesis tasks. Hence, $10$ ($52.6\%$) out of $19$ participants specified to have experience in the Computer Graphics area with $5$ of them being experienced in both Animation and Computer Graphics, $2$ ($10.5\%$) are working in the Biomechanics area, $1$ ($5.3\%$) participant is experienced only in Animation, and $6$ ($31.6\%$) are inexperienced. Out of $13$ participants being experienced in the above research areas, $1$ ($5.3\%$) individual claimed to have "Excellent" knowledge in their area, $6$ ($31.6\%$) claimed to have "Good" experience, and the remaining $7$ ($36.8\%$) have an "Average" level of experience in their area. The level of expertise was specified on a scale from "Excellent" to "Average", while individuals who do not have any experience in the mentioned research areas selected the "None" option.

In table \mbox{\ref{tab13}}, the percentage of participants selecting the fatigued motion sequence as the one they perceive as the most fatigued versus non-fatigued or less fatigued movements, is illustrated:

\begin{table}[ht]
\setlength\tabcolsep{2pt}
\centering
\begin{threeparttable}
\caption{Percentage of Participants Correctly Perceiving the Most Fatigued Motion}
{\begin{tabular}[t]{lccc}
\toprule
Motion & Animation Video Trial & Percentage & $\chi^2$ P-value \\
\midrule
Punching & NF VS $30\%$ F & $94.7\%$ & $0.0001$* \\
Waving & $30\%$ F VS $10\%$ F & $84.2\%$ & $0.0029$* \\
Throwing & $20\%$ F VS $40\%$ F & $73.7\%$ & $0.0389$*\\
Waving & $20\%$ F VS $40\%$ F & $89.5\%$ & $0.0006$* \\
Punching & NF VS $20\%$ F & $94.7\%$ &
$0.0001$* \\
Punching & $50\%$ F VS $30\%$ F  & $89.5\%$ & $0.0006$* \\
Throwing & $10\%$ F VS $50\%$ F  & $84.2\%$ & $0.0029$* \\
Punching & $40\%$ F VS NF & $89.5\%$ &
$0.0006$*\\
Throwing & $0\%$ F VS $30\%$ F  & $78.9\%$ &
$0.0116$*\\
Punching & NF VS $40\%$ F VS $20\%$ F & $73.7\%$ & $0.0389$* \\
Punching & $10\%$ F VS $20\%$ F VS $30\%$ F & $63.2\%$ & $0.1083$
\\
Throwing & $40\%$ F VS $20\%$ F VS $0\%$ F & $47.4\%$ & $0.4737$
\\
Punching & $0\%$ F VS $40\%$ F VS $20\%$ F &  $73.7\%$ & $0.0389$*
\\
Waving & $20\%$ F VS $0\%$ F VS $40\%$ F & $73.7\%$ & $0.0389$*\\
Throwing & $50\%$ F VS $0\%$ F VS $10\%$ F & $57.9\%$ & $0.2513$
\\
\bottomrule
\end{tabular}}
\label{tab13}
\begin{tablenotes}
\item[$^b$] {Various fatigued motion sequence comparisons were provided to the study participants in the form of animated videos via the questionnaire. The table illustrates the percentage of responders accurately perceiving the motion with the highest level of fatigue among others of the same motion type (either punching, throwing, or waving). In the last column, the $p$-values derived from the $\chi^2$ test are reported. $P$-values smaller than significance level $\alpha = 0.05$, are marked with an asterisk ($*$). "NF" and "F" stand for "Non-Fatigued and "Fatigued" motion, respectively.}
\end{tablenotes}
\end{threeparttable}
\end{table}%

As illustrated in Table \mbox{\ref{tab13}} (see column "Animation Video Trial"), we opted to create video sequences that depict both huge discrepancies in motion — such as a comparison between $10\%$ and $50\%$ fatigue-induced movements —as well as more subtle variations (e.g. motion sequences being $10\%$ fatigued versus $20\%$ fatigue). Furthermore, we randomized the order of the non/less/more fatigued motions in each animation trial, meaning that the most fatigued motion may appear on the right, left, or middle of the screen, to minimize participant bias. The participants were not provided with any additional information regarding the questionnaire, apart from its main aim; to perceptually evaluate fatigued human motion sequences.

Based on the findings from \mbox{\ref{tab13}} (refer to "Percentage" column), in the majority of cases, participants accurately perceived the animation trial that exhibited the most pronounced fatigue effects. As was expected, in cases where small differences in motion sequences are depicted, the percentage of participants correctly identifying the most fatigued motion is decreasing. For instance, in the trial where punching motion sequences of $10\%$, $20\%$, and $30\%$ of fatigue were compared, $12$ out of $19$ ($63.2\%$) participants detected the motion with the most fatigue effects. Furthermore, the throwing motion seems to have confused the participants the most, with only $9$ participants ($47.4\%$ of the study group) accurately pinpointing the most fatigued motion in an animated video containing motion sequences fatigued up to $20\%$ and $40\%$ as well as a non-fatigued throwing motion. One reason behind this is due to not providing the participants with multiple view angles of the same motion to aid them in detecting differences in the motions as the ones illustrated in Fig. \mbox{\ref{fig:7}} and those analyzed in Section \mbox{\ref{sec:fatigue_effects}}. Overall, asking responders to compare more than two fatigued motion sequences appeared to be challenging for them.

The participants were also asked to define the fatigue effects that they perceived as being conveyed through the fatigued animated sequences, for each motion type. Notably, this query was deliberately placed at the end of the questionnaire to prevent any potential bias, ensuring that participants did not consciously look for specific fatigue-related elements in the animated fatigued motions. The participants could choose more than one fatigue effect, while the option "Other" was added to the query to enable participants to add their own observations that made them choose the respective motion as the most fatigued one. Our findings regarding this query are outlined below:

\begin{itemize}
    \item \textit{Punching Motion}: Most of the participants were able to identify fatigue-induced elements within the fatigued motion sequences, such as i) the character's punches "missing" their target ($12$ - $63.2\%$ of participants), ii) the character being "unguarded" to hits ($9$ - $47.4\%$ of participants) and the character's arm performing a wider movement while hitting ($11$ - $57.9\%$ of participants). These observations are aligned with the results analyzed in Section \mbox{\ref{sec:fatigue_effects}}. On the contrary, some participants detected elements that are not present in the respective motions, such as the character's arm performing a more narrow movement while punching ($5$ - $26.3\%$ of participants). The same participants paired the latter observation with falsely assuming that the most fatigued motion sequence was the non-fatigued one in previous questions of the questionnaire.
    \item \textit{Throwing Motion}: As mentioned above, the participants struggled with both specifying the throwing animation trial that conveyed the most fatigued effects and the effects themselves. A small subset of the responders recognized some of the fatigue elements outlined in Section \mbox{\ref{sec:fatigue_effects}}, including the character throwing an invisible object to a different direction compared to the non-fatigued motion ($4$ - $21.1\%$ of participants), the character being inclined to the non-throwing side ($3$ - $15.8\%$ of participants) as well as the character's shoulder being positioned higher in the fatigued movement ($4$ - $21.1\%$ of participants). Moreover, the participants provided their own key observations, perceiving the character as being "out of balance" ($8$ - $42.1\%$ of participants), while unique responses were provided by individual participants ($5.3\%$ of participants) highlighting that the character's arm appeared "more tired and lower (placed) as it should" and "less range of motion, failure to perform the whip-like movement", was noted. One responder stated that "all looked exactly the same", stressing their inability to distinguish between the different levels of fatigue within the throwing motion animated videos.
    \item \textit{Waving}: For the waving motion, the majority of responders ($13$ - $68.4\%$ of the participants) were able to discern that the waving hand exhibits a limited range of motion in the fatigued motion, while $11$ participants ($57.9\%$) perceived that "the character struggles to perform the task".   
\end{itemize}

Using a five-point Likert scale - where $1$ corresponds to "Very Unrealistic" and $5$ to "Very Realistic" — participants were asked to evaluate the extent to which the fatigued motions they were exposed to convey the externally perceived fatigue. \comm{\textcolor{red}{The assessment of fatigued motion sequences across the motion types of punching, throwing, and waving in terms of realism is depicted in Fig. .....}} The majority of responders characterized the fatigued-induced motion that our Fatigue-PINN model produces as "realistic" across all motion types. In particular, $12$ participants ($63.2\%$) found the fatigued punching motion "realistic", while the rest of the results were formed as follows: $2$ ($10.5\%$) participants found the fatigued punching motion as "unrealistic", $4$ ($21.1\%$) as "neutral" and $1$ ($5.2\%$) as "very realistic".

Nine ($47.4\%$) of the participants described the throwing motion as "realistic", while $6$ ($31.6\%$) responders characterized it as "Neutral", being in line with their responses in previous questions (i.e. not differentiating between fatigued and non-fatigued throwing motions). The realism ratings for the fatigued throwing motion sequences are complemented with $3$ ($15.8\%$) participants describing them as "unrealistic" and $1$ ($5.2\%$) as "very realistic".  No "very unrealistic" ratings were reported for the fatigued punching and throwing motions. The fatigued waving motion was also described as "realistic" by more than half of the study group ($10$ participants - $52.6\%$), while $3$ ($15.8\%$) participants commented on the motion being "very realistic". On the contrary, another $3$ ($15.8\%$) responders found the fatigued waving movements "unrealistic", $1$ ($5.2\%$)"very unrealistic", and $2$ ($10.5\%$) as "neutral".

We also requested the feedback of the participants regarding what fatigue elements they expect to see from either a punching, throwing, or waving fatigued motion. In the case of a fatigued punching motion sequence, some of the participants ($8$ - $42.1\%$) commented on fatigue elements already present in our produced animations, such as "lower positioning" and "wider movement" of the hands, as well as "less consistent moves" (i.e. "missing target", "struggling to hold up arms"), while more than the half of the study group ($10$ - $52.6\%$) were expecting to see "slower motion". Other interesting remarks consist of "bad balance, bad posture", "mechanical representation of weight (hunched shoulders, bad posture, heaving)" as well as "legs should also display more fatigue, like less or more restricted movement". Thus, we conclude that the participants were not able to discern alterations in the character's posture such as in lumbar bending and rotation rendering the character bending more to the left side while hitting with the right hand, or in the torso's rotation resulting in turning more to the right, as described in Section \mbox{\ref{sec:fatigue_effects}}. Furthermore, it was expected for the responders to observe the absence of fatigue in the legs, since in the context of the movements investigated in this work, we opted to apply fatigue only on the upper body.

Regarding the fatigued throwing motion, similar comments to the ones received for the punching one, were collected from the participants, including observations from the motion produced via Fatigue-PINN ("wider motion of the hands", "hands being placed lower", "leaning", "discrepancies in the position and rotation of the shoulder and elbow to fulfill the motion" etc.) and other effects of externally perceived fatigue (e.g. "bad balance, maybe slipping", "limited range of motion", "fatigue also in legs", "lacking full arm extension"). The presence of fatigue in the character's torso was also highlighted by the participants, but they anticipated to perceive more intense fatigue effects during throwing movements (e.g. "I would expect that the character would use its torso more instead of the arm.").
As for the waving motion, the participants reported expecting to see the following fatigue elements in a fatigued waving motion: "elbow flexion featuring a narrow range of motion", (similarly) "waving angle to be shorter", "difficulty to wave", "a slighter raise of the hand" and "bad posture". However, $2$ ($10.5\%$) of the participants, were not able to determine such fatigue effects in waving motion "without context", and commented that they perceive the animated results depicted in the videos within the questionnaire to be evident in cases where the levels of fatigue are relatively high ("less proclaimed fatigue in general at easier tasks like waving"). In all motion types, $1$ or $2$ participants were expecting to also see different "facial/sad expressions" to complement the overall sentiment of fatigue, which, however, is out of the scope of this work.

We hypothesize that by providing participants with longer-in-duration animated videos as well as videos of the same motion at different frame speeds ("slow-motion" videos, etc.), the participants may be able to detect more fatigued-induced characteristics in the respective motions, not "missing" discrepancies in the motion of the torso and shoulders.

As for statistical analyses, we performed a chi-square ($\chi^2$) Test of Independence at significance level $\alpha = 0.05$, to examine whether there is a significant association between the participants' selections of the most fatigued motion sequences and the expected results (“actual” fatigued motions). The null hypothesis states that no such relationship exists, and a $p$-value smaller than $\alpha$ indicates the rejection of the null hypothesis. Observing Table \mbox{\ref{tab13}} (see column "$\chi^2$ P-value") in $12$ out of $15$ tests, $p < 0.05$, meaning that the null hypothesis is rejected, hence participants opted for the most fatigued motion sequence accurately (i.e. there is a significant correlation between the participants' perception of fatigue across fatigue levels and the "correct" answers). The cases where $p > 0.05$ were the ones where participants struggled to correctly discern fatigue differences in the animated motions, resulting in non-statistically significant outcomes. 

Finally, a One-Way Analysis of Variance (ANOVA) was utilized to determine whether the different fatigue levels illustrated in the animated videos provided to the participants led to significantly different realism ratings obtained through Likert scales. The null hypothesis in this case states that no statistically significant differences in realism ratings are present, i.e. responders gave consistent realism evaluations, which is a desirable outcome. As can be deduced from Table \mbox{\ref{tab14}}, all $p$-values $>= 0.05$, in which case the null hypothesis is not rejected, meaning that the realism perception of the study participants does not differ significantly while assessing motion sequences conveying different levels of fatigue. Therefore, the participants consistently characterizing the fatigued motions as "realistic" is a statistically significant result.

\begin{table}[ht]
\setlength\tabcolsep{5pt}
\centering
\begin{threeparttable}
\caption{ANOVA $P$-values for Fatigue Level Realism Evaluations}
{\begin{tabular}[t]{lccc}
\toprule
 & Punching Motion & Throwing Motion & Waving Motion \\
\midrule
$p$-value & $0.1782$ & $0.3630$ & $0.4254$\\
\bottomrule
\end{tabular}}
\label{tab14}
\begin{tablenotes}
\item[$^c$] {The $p$-values as arisen by One-Way ANOVA tests performed to specify whether the realism perception of participants while assessing fatigued motion sequences differs significantly across the fatigue levels of the motions.}
\end{tablenotes}
\end{threeparttable}
\end{table}%

Both statistical analyses were implemented using built-in Python functions.

\subsection{Accuracy Results}

In Table \ref{tab3} we provide the Normalised Root Mean Squared Error (NRMSE) values for joint-specific ID and FD models of the Fatigue-PINN framework across the motion trials of an action. These NRMSE values are average values computed from the results of the three validation folds for each motion type, namely waving, punching, and throwing. \comm{NRMSE is a scale-invariant measure given by Eq.(\ref{eq:8}):
\begin{equation}
NRMSE = \frac{\sqrt{MSE}}{max(\textbf{y}) - min(\textbf{y})} * 100\%
\label{eq:8}
\end{equation}
\noindent where $max(\textbf{y})$ and $min(\textbf{y})$ are the maximum and minimum values across all time points for the ground-truth joint angle/torque data. Therefore, }NRMSE values indicate the discrepancies between ground-truth and predicted joint angles/torques during the testing of the models, thus, smaller values signify good accuracy. As observed in Table \ref{tab3} all values are $<1$, consequently, our models perform quite well in new (unseen) data.

\begin{table}[ht]
\setlength\tabcolsep{1.5pt}
\centering
\begin{threeparttable}
\caption{NRMSE For ID/FD Models}
{\begin{tabular}[t]{lcccccc}
\toprule
\multicolumn{7}{c}{NRMSE values For Right Upper Limb Joint Torques/Angles} \\
\midrule
Motion & arm\_flex & arm\_add & arm\_rot  & elbow\_flex & elbow\_pro & wrist\_flex \\
\midrule
Punching & $0.17$/$0.72$ & $0.11$/$0.90$ & $0.91$/$0.36$ & $0.75$/$0.13$ & $0.16$/$0.37$ & $0.27$/$0.99$ \\
Throwing & $0.49$/$0.58$ & $0.12$/$0.57$ & $0.28$/$0.69$ & $0.39$/$0.62$ & $0.14$/$0.56$ & $0.19$/$0.74$  \\
Waving & $0.21$/$0.24$ & $0.23$/$0.24$ & $0.11$/$0.72$ & $0.31$/$0.25$ & $0.93$/$0.19$ & $0.58$/$0.14$\\
\midrule
\multicolumn{7}{c}{NRMSE values For Right Lower Limb Joint Torques/Angles} \\
 \midrule
 Motion & hip\_flex & hip\_add & hip\_rot  & knee\_angle & ankle\_angle & sub\_angle\\
\midrule
Punching & $0.19$/$0.20$ & $0.23$/$0.79$ & $0.37$/$0.55$ & $0.17$/$0.28$ & $0.45$/$0.71$ & $0.57$/$0.63$ \\
Throwing & $0.10$/$0.57$ & $0.20$/$0.62$ & $0.39$/$0.65$ & $0.73$/$0.50$ & $0.80$/$.79$ & $0.13$/$0.67$\\
Waving & $0.35$/$0.32$ & $0.41$/$0.15$ & $0.10$/$0.18$ & $0.40$/$0.20$ & $0.27$/$0.12$ & $0.34$/$0.25$  \\
\midrule
\multicolumn{7}{c}{NRMSE values For Joint Torques/Angles of the Torso} \\
\midrule
 Motion & pelv\_tilt & pelv\_list & pelv\_rot  & lumb\_ext & lumb\_bend & lumb\_rot \\
\midrule
Punching & $0.22$/$0.15$ & $0.29$/$0.28$ & $0.35$/$0.30$ & $0.16$/$0.31$ & $0.25$/$0.24$ & $0.16$/$0.32$ \\
Throwing & $0.99$/$0.12$ & $0.19$/$0.57$ & $0.71$/$0.44$ & $0.38$/$0.58$ & $0.66$/$0.71$ & $0.52$/$0.58$  \\
Waving & $0.37$/$0.20$ & $0.36$/$0.30$ & $0.19$/$0.15$ & $0.20$/$0.29$ & $0.30$/$0.39$ & $0.59$/$0.64$  \\
\bottomrule
\end{tabular}}
\label{tab3}
\begin{tablenotes}
\item[$^d$] {NRMSE  values for joint torques/angles of the right upper and lower limb as well as torso, for punching, throwing, and waving motions. These NRMSE values are derived by averaging the prediction results across all validation folds of each action. The first values in each row indicate the NRMSE values produced by the joint-specific ID models, while the second values (after the "/") denote the NRMSE values for joint-specific FD models. \comm{The NRSME metric is unitless as it arises from Eq. (\ref{eq:8}) ($\frac{\sqrt{N^2}}{N}*100\%$).}}
\end{tablenotes}
\end{threeparttable}
\end{table}%

To further support the selection of a joint-specific approach, i.e. apply Fatigue-PINN for each joint angle, against the non-joint-specific, we provide the Pearson's Correlation Coefficient, $R^2$, values to evaluate the precision of the two methods. $R^2$ evaluates the correlation between the predicted joint torques/angles and their real values, with values closer to $1$ indicating better alignment. As shown from Table \ref{tab4} the predictions of the ID and FD models of the joint-specific architecture tend to be more correlated (higher $R^2$) with the ground-truth values, than the ones of the non-joint-specific approach, providing an insight into average testing accuracy for the former method being $\sim97\%$, in contrast of average testing precision for the non-joint-specific, $\sim86\%$. As expected, the joint-specific Fatigue-PINN architecture enhances the precision of animations by leveraging neural networks to extract latent space correlations from all joint torques/angles in order to efficiently accomplish a single task.
It is worth noting that the $R^2$ values are computed for the joint torques/angles of the upper right limb and during punching motion. Akin to the computation of the NRMSE values, the $R^2$ values are also average values calculated from the predictions of the three validation folds of an action (in our case the punching motion). Similar results with minimum differences $\sim1-2\%$ are observed in the other two movements, as well as the rest of the joints. 

\vspace{-5pt}
\begin{table}[ht]
\setlength\tabcolsep{1.5pt}
\centering
\begin{threeparttable}
\caption{$R^2$ For Joint-Specific and Non-Specific Approaches}
{\begin{tabular}[t]{lcccccc}
\toprule
\multicolumn{7}{c}{$R^2$ values For Right Upper Limb Joint Torques/Angles during Punching} \\
\midrule
Model & arm\_flex & arm\_add & arm\_rot  & elbow\_flex & elbow\_pro & wrist\_flex  \\
\midrule
JS Model & $0.97$/$0.96$ & $0.90$/$0.91$ & $0.96$/$0.95$ & $0.94$/$0.96$ & $0.93$/$0.97$ & $0.94$/$0.95$  \\
NJS Model & $0.88$/$0.90$ & $0.81$/$0.85$ & $0.84$/$0.88$ & $0.85$/$0.87$ & $0.82$/$0.89$ & $0.81$/$0.82$  \\
\bottomrule
\end{tabular}}
\label{tab4}
\begin{tablenotes}
\item[$^e$] {Average (across all validation folds of a motion type) Pearson's Correlation Coefficient values ($R^2$) for predictions of ID (estimating joint torques) and FD (estimating joint angles) models, during punching motion. A comparison of $R^2$ values between the Fatigue-PINN joint-specific approach and the non-joint-specific approach is provided. Thus, JS stands for "Joint-Specific", and NJS, for "Non-Joint-Specific" models. Just as in \ref{tab3}, $R^2$ values of the ID models are presented first, while the second values (after the "/") denote the $R^2$ values for the respective FD models.}
\end{tablenotes}
\end{threeparttable}
\end{table}

In order to validate quantitatively the visual results discussed in Section \mbox{\ref{sec:fatigue_effects}}, we compute the $R^2$, and L2 metrics to measure the discrepancies in joint angles before and after applying a certain level of fatigue on each one of the different motion types (punching, throwing and waving). The values of these metrics are reported in Table \mbox{\ref{tab15}}, for the shoulder flexion, shoulder adduction, elbow flexion, lumbar bending, and lumbar rotation joint angles, whose fatigue-induced alterations in motion are mostly discussed in Section \mbox{\ref{sec:fatigue_effects}}. L2 provides a quantification of the "distance" between joint angles before and after injecting fatigue in motion. As shown in Table \mbox{\ref{tab15}}, increases in the level of fatigue imply higher L2 values, meaning greater differences in joint angles, and hence more profound visual fatigue effects. The latter observation is valid for all types of motion, while $R^2$ values remain relatively high, indicating that there is still a correlation between the non-fatigued and fatigued data. It is worth noting that some angles, like shoulder flexion, are more affected by fatigue compared to others. Concluding, the throwing motion exhibited smaller differences in joint angles across fatigue levels compared to the other two motions, justifying why the study participants were not able to perceive evident changes due to fatigue in throwing motion sequences.

\begin{table}[ht]
\setlength\tabcolsep{1.5pt}
\centering
\begin{threeparttable}
\caption{L2/$R^2$ values For Right Upper Limb Joint Angles}
{\begin{tabular}[t]{lccccc}
\toprule
\multicolumn{6}{c}{Punching Motion} \\
\midrule
Motion & arm\_flex & arm\_add & elbow\_flex & lumbar\_bend & lumbar\_rot \\
\midrule
NF VS $20\%$ F & $2.01$/$0.92$ & $2.71$/$0.90$ & $1.98$/$0.92$ & $1.36$/$0.94$ & $2.63$/$0.90$ \\
NF VS $30\%$ F & $3.21$/$0.90$ & $4.24$/$0.88$ & $2.07$/$0.92$ & $2.12$/$0.90$ & $3.33$/$0.89$\\
NF VS $40\%$ F & $4.37$/$0.89$ & $5.44$/$0.87$ & $2.40$/$0.90$ & $3.07$/$0.91$ & $4.02$/$0.90$\\
NF VS $50\%$ F & $5.12$/$0.88$ & $6.23$/$0.85$ & $2.59$/$0.91$ & $3.49$/$0.90$ & $4.38$/$0.89$ \\
\midrule
\multicolumn{6}{c}{Throwing Motion} \\
\midrule
Motion & arm\_flex & arm\_add & elbow\_flex & lumbar\_bend & lumbar\_rot \\
\midrule
NF VS $20\%$ F & $1.08$/$0.96$ & $1.07$/$0.94$ & $1.62$/$0.93$ & $0.93$/$0.95$ & $1.04$/$0.94$ \\
NF VS $30\%$ F & $1.50$/$0.95$ & $1.58$/$0.93$ & $1.81$/$0.93$ & $1.14$/$0.94$ & $1.20$/$0.94$ \\
NF VS $40\%$ F & $1.95$/$0.93$ & $2.34$/$0.92$ & $2.44$/$0.92$ & $1.42$/$0.92$ & $1.39$/$0.93$ \\
NF VS $50\%$ F & $2.40$/$0.90$ & $2.75$/$0.91$ & $2.49$/$0.89$ & $1.86$/$0.92$ & $1.93$/$0.90$ \\
\midrule
\multicolumn{6}{c}{Waving Motion} \\
\midrule
Motion & arm\_flex & arm\_add & elbow\_flex & lumbar\_bend & lumbar\_rot \\
\midrule
NF VS $20\%$ F & $1.29$/$0.96$ & $0.93$/$0.97$ & $1.18$/$0.94$ & $1.30$/$0.95$ & $1.45$/$0.93$ \\
NF VS $30\%$ F & $1.52$/$0.96$ & $1.39$/$0.96$ & $2.02$/$0.91$ & $1.91$/$0.95$ & $2.34$/$0.92$ \\
NF VS $40\%$ F & $1.81$/$0.96$ & $2.16$/$0.95$ & $2.87$/$0.91$ & $2.66$/$0.94$ & $3.26$/$0.90$ \\
NF VS $50\%$ F & $2.16$/$0.94$ & $2.66$/$0.94$ & $3.00$/$0.91$ & $3.23$/$0.92$ & $3.53$/$0.91$ \\
\bottomrule
\end{tabular}}
\label{tab15}
\begin{tablenotes}
\item[$^f$] {L2/$R^2$ metrics measured between the values of upper body joint angles before and after applying a certain level of fatigue on motion. The first values in each row indicate the L2 values while the second values (after the "/") denote the $R^2$ values. "NF" and "F" stand for "Non-Fatigued and "Fatigued" motion, respectively.}
\end{tablenotes}
\end{threeparttable}
\end{table}%


\vspace{-7pt}
\section{Discussion}

\subsection{Ablations}
\label{seq:Ablations}

Within Fatigue-PINN, we modeled our ID and FD networks as PINNs as well. The primary distinction in our corresponding BiLSTM networks lies in the loss function, where, in addition to the MSE term, which minimizes the loss between the neural network estimations and the real values of joint actuator torques (for ID) and joint angles (for FD), we incorporated a physics-based loss term derived from the classical equation of motion (Eq. (\ref{eq:L_ID})). This equation denotes the relationship of joint kinematics (generalized positions, i.e. joint angles, velocities, and accelerations), with generalized forces (i.e. joint torques) during human motion: 

\begin{equation}
    L_{{PB}_{ID}} = \frac{1}{N}\sum_{n=1}^{N}\sum_{t=0}^{T}(M(\mathbf{q}_t^n)\ddot{\mathbf{q}}_t^n + C(\mathbf{q}_t^n, \dot{\mathbf{q}}_t^n) + G(\mathbf{q}_t^n) - \mathbf{\tau}_t^n)^2 
    \label{eq:L_ID}
\end{equation}

where, $N$ is the number of DoFs, $M(\mathbf{q}_t^n) \in R^{NxN}$ is the Mass matrix, $C \in R^N$ is the Coriolis and Centrifugal forces vector and $G \in R^N$ is the Gravitational forces vector, while $\mathbf{q},\mathbf{\dot{q}},\mathbf{\ddot{q}} \in R^N$ are the vectors of generalized positions, velocities, and accelerations respectively, and $\mathbf{\tau}_t^n$ represent joint torques. 

Nevertheless, the changes in model accuracy and generated fatigued motion are insignificant whether we use PINNs to formulate our ID and FD models or not (the NRMSE values of the ID/FD PINN models differ by about $\pm 5\%$ compared to those in Table \ref{tab3}). Therefore, even though we utilized the standard BiLSTM architecture for the ID and FD models, their PINN versions offer a viable solution in cases of \comm{limited data availability}few training data. This is because PINN loss functions act as soft physical constraints during human motion, facilitating the inference of corresponding joint torques and angles.

It is worth mentioning that we also experimented with training the 3CC-$\lambda$ unsupervisingly (solving the forward problem), where no ground-truth data for fatigued, $M_F$, and resting state $M_R$, are provided. 
In this case, the PINN loss, $L$ (see Eq. (\ref{eq:3CC-loss1})), still has its physics-based term, $L_{PB}$ (Eq. (\ref{eq:3CC-loss3})), and instead of $L_{NN}$ (Eq. (\ref{eq:3CC-loss2})), we add a loss term to incorporate boundary conditions regarding $M_F$ and $M_R$. To evaluate which of the two training schemes, supervised or unsupervised, is more efficient for the 3CC-$\lambda$ model, we calculated the Pearson's Correlation Coefficient, $R^2$, between predicted and ground-truth (experimental) values. In the unsupervised scheme, the $R^2$ for each joint angle across emotional trials of an action (punching, waving, or throwing) range from $\sim0.93-0.96$, while the supervised version of the model has $R^2$ values of $\sim0.91-0.97$. Hence, the two approaches provide 3CC-$\lambda$ models whose $R^2$ values differ by $\sim0.01-0.03$ units. This small difference in accuracy between the two 3CC-$\lambda$ versions is probably due to the fact that the equations driving the 3CC are relatively simple, facilitating the convergence to the correct solution, even in the unsupervised version. The latter enables us to follow either of the two training processes for our 3CC-$\lambda$ model, however, training 3CC-$\lambda$ unsupervisingly requires more computational resources compared to the approach outlined in Section \ref{sec:3CC}.  


\subsection{Fatigued Movements}






Fatigue is usually neglected in most musculoskeletal models as well as in data-driven motion synthesis frameworks, leading 
in bypassing an element that could significantly boost animation realism.
The widely emerging Physics-Informed Neural Networks are also making their debut in human biomechanics estimation tasks (e.g. prediction of joint angles/forces/moments, muscle forces, etc. \cite{Zhang2023, Ma2024, ZhiboZhang2022, Taneja2024, Kumar2023, Zhang2022_2, Kumar2024, Zhang2023_2}) since they can \comm{inject physics-based domain knowledge into the learning procedure to both approximate PDEs describing the problem-to-solve (e.g. Inverse of Forward Kinematics equations) and penalize solutions not adhering to physical laws, limiting the solution space, which leads to faster convergence during training. Consequently, PINNs}provide more accurate physically consistent solutions for both forward and inverse problems, while requiring a significantly smaller amount of data to approximate the underlying physical laws. 

In this study, we present Fatigue-PINN, an advanced end-to-end PINN-based deep learning framework to synthesize fatigued human motion without relying on fatigued motion capture data for training. Our architecture is composed of three modules, the Inverse Dynamics Module, which translates input joint angles into joint torques, the Fatigue module, i.e. the 3CC-$\lambda$ model to simulate
the effects of fatigue in the maximum exerted joint torques, and the Forward Dynamics model, which transforms these fatigued joint torques back into joint angles, thus enabling the fatigued animation of a 3D character. 3CC-$\lambda$ is a PINN adaptation of the 3CC state machine \cite{Liu2002, Xia2008, Frey-Law2012}, augmented with a method to account for joint-specific fatigue configurations, thereby ensuring seamless incorporation of fatigue effects in different motions and avoid sharp step-wise motions. Additionally, we use ID and FD models employed upon each DoF. These joint-specific models enhance predictive accuracy and are also compatible with the joint-specific fatigue profiles as defined by 3CC-$\lambda$. By having control at the joint level, our model is capable of generating physically plausible movements with reduced motion artifacts.
The ID and FD models are LSTM-based to extract the temporally evolving fatigue impact and can be utilized as surrogate models to perform the respective tasks without the need for GRFs. GRFs are expected to vary between non-fatigued movements and fatigued movements, to account for changes in the body’s center of mass/pressure, due to variations in posture and kinematic patterns \cite{Fuller2009, Cogswell2016}. In simpler words, alterations in upper limb torques (the ones investigated in this work) due to fatigue, will influence GRFs. Thus, in order to include GRFs in the training/inference of our ID and FD models, the use of training datasets that contain both fatigued motion capture and GRF data, is required. Such datasets with motion sequences covering open-type movements, as the ones explored in this work, are not available in the literature. 

Our results indicate that Fatigue-PINN produces \comm{realistic} fatigue effects on open-type movements quite similar to those of externally perceived fatigue, which are consistent with the existing literature \cite{Fuller2009, Dunn2017, Dunn2019, Haralabidis2020, Yang2019}. 
It is worth mentioning that in our experiments, at each action (i.e. waving, punching, and throwing), all upper body muscles were fatigued isometrically, causing overlapping kinematic changes to render the externally perceived fatigue impact more evident. For instance, for the punching motion, the $\lambda$ factor for all upper body joints is set to $0.6$, which provides a good balance between motion smoothness and conveyance of externally perceived fatigue. Nevertheless, since our Fatigue-PINN framework is joint-specific it can also be utilized to study the effects of fatigue on posture and joint kinematics after inducing localized muscular fatigue (e.g. fatiguing the shoulder and observing the impact on elbow and trunk), just as in \cite{Yang2019}, where no data-driven method was used.  

A perceptual study was also conducted indicating that the participants were able to distinguish between the various levels of fatigue of a motion and further characterize which fatigued motion exhibited the most pronounced fatigue effects. Moreover, the participants succeeded in specifying the latter effects in fatigued motion sequences across punching, throwing, and waving motion types, as well as providing their own visual observations, pinpointing fatigued effects that the authors were incapable of defining. However, it is worth mentioning that fatigued throwing movements, posed a challenge to the responders, rendering them unable to achieve high percentages of accurate distinctions between fatigued and non-fatigued throwing motions. Despite this fact, the majority of participants described the fatigued motion sequences of all motion types (punching, throwing, and waving) as “realistic”. Moreover, we assume that providing the participants with longer-in-duration animated videos, at different frame speeds, as well as featuring multiple view angles of the same motion, would aid the responders to perceive more fatigued-induced elements in the respective motions. The outcome of the perceptual study was further supported by a statistical analysis based on $\chi^2$ and One-Way ANOVA tests.

Furthermore, our end-to-end Fatigue-PINN framework can be utilized to both facilitate the work of animators and aid biomedical engineers with investigating the effects of externally perceived fatigue on posture and movements. By simply providing a motion sequence (i.e. a sequence of joint angles), the fatigued movement at a 
specified level of fatigue is synthesized. Fatigue-PINN does not require direct input of joint torques for fatigue incorporation, as it can convert joint angles into torques and vice versa. The latter enables the seamless integration into motion synthesis pipelines that operate on joint angles.

A limitation of our work is that even though the prediction metrics of the Fatigue-PINN framework indicate a good accuracy in closed-type movements, i.e. object manipulation or multi-contact motions, this is not reflected in the animation results of such motions, with motion artifacts being present, mostly in the end effectors. More specifically, we evaluated the Fatigue-PINN framework on the "lifting a box" and "pushing a box" motion types from the Adapt Emotional Actions dataset. In both movements, the overall testing accuracy of Fatigue-PINN is quite high, around $92-95\%$, while the average -across motion trials- NRMSE and $R^2$ values of ID/FD models for right upper limb joint torques/angles are very close to those reported in Tables \mbox{\ref{tab3}} \& \mbox{\ref{tab4}}. As for the produced fatigue animation, since our framework does not support contact integration, in both box manipulation motions, the hands are misplaced, resulting in either penetrating or not touching the “ghost” of the box. Also, foot skating artifacts are present. A possible solution to fix hand artifacts is to incorporate in our framework contact points \cite{Starke2019, Starke2020} as well as forces produced by human-object interactions. As for foot contact artifacts, exploiting GRFs and foot contact information  \cite{Mourot2022} as well as utilizing GANs \cite{Wang2019} or diffusion models \cite{Raab2023}, are possible ways to mitigate these artifacts. The latter is proposed as future work. 

Another limitation is that the dataset utilized in this work is recorded upon the expressive movements of one subject. Thus, even though the diverse motion trials improve the generalization ability of our ID and FD models across variations of the same motion (i.e. the same movement expressed via different emotions), the absence of subject variability hinders the evaluation of our model in terms of generalization across subjects. In addition, a dataset containing more motion trials for each action movement would be preferable to further enhance the robustness of the model. More experiments on larger datasets with both subject and motion trial diversity remain as future work. 

Finally, the Adapt Emotional Actions dataset does not provide any fatigued motion capture data, and, as far as the authors are concerned, fatigued motion capture datasets covering open-type movements as the ones utilized in this work, are not present in the literature or open-source. 
The inability to validate our framework against real fatigued motion capture data is also a limitation of our work, and we are currently working on an expansion of Fatigue-PINN, incorporating fatigue in the lower limbs, such as gait movements, where actual fatigued data are available through numerous publications. In the latter case, the comparison between our framework and baseline methods will be facilitated.

\vspace{-7pt}
\section{Conclusion}

Addressing the gap of data-driven methods for modeling fatigue in literature, we introduce Fatigue-PINN, a PINN-based deep learning framework for synthesizing fatigued human movements without explicitly modeling the complex underlying physics laws governing the human body. Our framework consists of three modules, namely Inverse Dynamics, Fatigue, and Forward Dynamics modules, forming an encoder-decoder-like architecture for producing fatigued joint angle sequences based on fatigued-induced joint torques. Despite the incorporation of fatigue within the torque space, the end-to-end encoder-decoder configuration processes joint angles to produce their fatigued adaptations, facilitating the integration with any motion synthesis model functioning on joint angles and testing the real-time capabilities of the system. To estimate the fatigued state of motion, we model the 3CC state machine as a PINN, 3CC-$\lambda$, and we enhance it with the parameter $\lambda$ to characterize the fatigue profile of each joint and contribute to the smooth integration of fatigue in produced motion, thus, preventing abrupt frame transitions. Our semi-dynamics joint-specific approach, employing an ID and FD BiLSTM model for each joint, not only improves the overall predictive accuracy but also enables better alignment with joint-specific fatigue configurations and handling motion artifacts. Fatigue-PINN simulates the effects of externally perceived fatigue on open-type human movements comparable with findings from real-world experimental fatigue studies, even though it does not require fatigued motion capture data for the learning process. The adaptation of Fatigue-PINN to closed-type movements involving contacts and forces in human-object interactions as well as periodical movements where lower limbs are mostly involved, like gait, remains a future research direction. 


\appendix
\section{\break Pipeline Details}

\begin{figure*}[h!]
\centering
\includegraphics[width=\textwidth]{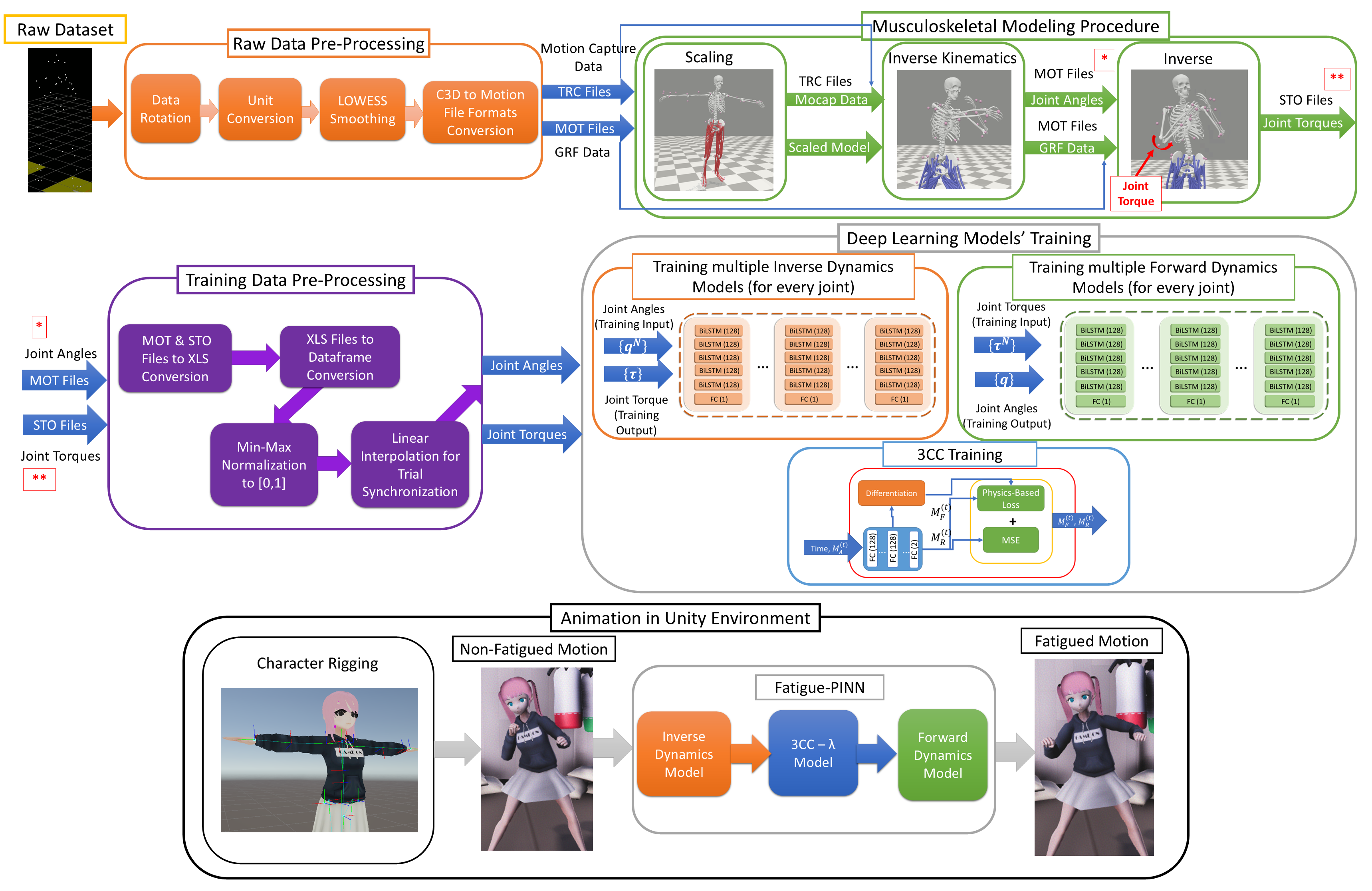}
\caption{An overview of our pipeline to develop and test the Fatigue-PINN framework.}\label{fig:11}
\end{figure*}

The aim of this Appendix is to provide more insights regarding the procedure we followed to develop and train our Fatigue-PINN framework, in order to aid researchers with reproducing our results. The full Fatigue-PINN pipeline consists of the following modules:
\begin{enumerate}
    \item Raw Data Pre-Processing
    \item Musculoskeletal Modeling 
    \item Training Data Pre-Processing
    \item Deep Learning Models' Training
    \item Animation in Unity Environment
\end{enumerate}
as illustrated in Fig. \ref{fig:11}. 

\subsection{Raw Data Curation}
\label{sec:raw_data_curation}

The first step of our pipeline is to pre-process the raw motion capture and GRF data contained in the Adapt Emotional Actions open dataset, utilized for the purposes of this work. Since these data are in C3D format, we created a Python script that converts C3D files to motion file formats that the OpenSim musculoskeletal modeling software can process. In particular, these scripts make use of OpenSim Python API's built-in functions to read C3D files and write marker data in TRC files and their respective force plate data in MOT files. By also using functions from the OpenSim Python library, marker, and GRF data are rotated over $-90$ degrees along the x-axis and $270$ degrees over the y-axis to match the dataset's experimental coordinate system to the one of OpenSim. The raw data of each trial of every motion type investigated in this work is saved in individual motion files.

Before saving, GRF data are also converted from millimeters to meters to be aligned with the measurement units of OpenSim and smoothed via a Locally Weighted Regression algorithm. This method, also known as Locally Weighted Scatterplot Smoothing (LOWESS), fits a non-parametric regression curve to a scattered set of data. More specifically, unlike traditional models that apply a single global model, LOWESS fits multiple models over local regions of data, weighing nearby points higher based on their proximity to the target value (the value where we want to estimate the output). Thus, LOWESS flexibly adapts to complex datasets, handling noisy and non-linear data. In our implementation of LOWESS, we utilized a bell-shaped kernel function with parameter $c$, where larger $c$ values will result in a smoother curve. The aforementioned process, i.e. measurement unit conversion and smoothing,  is implemented through Python scripting. 

\comm{
\begin{algorithm}
\caption{Locally Weighted Regression with Bell-Shaped Kernel}
\begin{algorithmic}[1]
\Require $t_{np}$ (time array), $v_{int}$ (value matrix), $\tau$ (smoothing parameter)
\Ensure $vest$ (smoothed values)

\State Initializing all weights from the bell shape kernel function
\For{$j = 1$ to $3$}
    \For{$i = 1$ to $r$}
        \State $w[i] \gets \exp(- (t_{np} - t_{np}[i])^2 / (2\tau))$
    \EndFor
\EndFor

\State Looping through all v-points 
\For{$j = 1$ to $3$}
    \For{$i = 1$ to $r$}
        \State $weights \gets w[i]$
        \State Compute $b = [\sum (weights \cdot v_{int}[:, j]), \sum (weights \cdot v_{int}[:, j] \cdot t_{np})]$
        \State Compute matrix $A$ from weighted sums
        A = \begin{bmatrix} \sum weights & \sum (weights \cdot t_{np}) \\ \sum (weights \cdot t_{np}) & \sum (weights \cdot t_{np} \cdot t_{np}) \end{bmatrix}
        \State Solve $A \theta = b$ for $\theta$
        \State $vest[i, j] \gets \theta[0] + \theta[1] \cdot t_{np}[i]$
    \EndFor
\EndFor

\Return $vest$
\end{algorithmic}
\end{algorithm}}

\subsection{Musculoskeletal Modeling Procedure}

We used the OpenSim \cite{Seth2018} open-source tool to derive the joint kinematics and dynamics (especially joint angles and torques), based on the markers’ spatial trajectories and GRFs that were processed through the above procedure (see Section \ref{sec:raw_data_curation}). As depicted in Fig. \ref{fig:11}, the musculoskeletal modeling procedure includes the following analyses:
\begin{itemize}
    \item Scaling: Motion capture data containing the subject’s neutral pose marker data are used to scale a generic musculoskeletal model (that is, a Hill-type full-body presented in \cite{Rajagopal2016}), to the anthropometric measurements of the respective subject (e.g., length of each body segment, height, etc.). Hence, obtain a personalized model of the subject whose motion was recorded for the Adapt Emotional Actions dataset. In order to scale the dimensions of each body segment, the Scaling tool ensures that the distances between the virtual markers of the OpenSim model match the distances between the experimental markers (the ones from motion capture). To optimize the scaling process, the virtual markers' positions were manually modified before applying the scaling tool. The output of this step is the scaled model. 
    \item Inverse Kinematics: The IK tool is employed upon the scaled model to reproduce the recorded motion from the experimental markers' spatial trajectories (motion capture data). In particular, at each time frame, the error between experimental (motion capture) markers and musculoskeletal model markers' position is calculated by IK to estimate the best matching pose. The latter process is carried out via a Weighted Least Squares equation, where a sum of marker weighted squared errors is minimized. Through the IK process, the generalized positions, i.e., joint angles of the whole body are extracted and stored in a MOT file format. 
    \item Inverse Dynamics: The joint angles computed via IK and the experimentally measured GRFs are used as an input to the Inverse Dynamics tool to calculate the generalized forces, i.e. joint torques, as mandated by Eq. (\ref{eq:L_ID}). Joint torques are stored in a motion file of STO format. 
\end{itemize}

The inputs and parameters for each analysis (e.g. input files, duration of the motion to be analyzed, etc.) are specified through XML files, automating the OpenSim pipeline. 

\subsection{Training Data Pre-Processing}

Following the musculoskeletal modeling process, joint kinematics and torques are extracted from experimental motion capture and force plate data. Before using this derivate data for the training of our ID/FD BiLSTM models, further processing is required. The processes outlined in this section are conducted using Python scripting since the models that constitute our Fatigue-PINN framework were trained in Python Keras. 

The motion files containing generalized positions (MOT files) and forces (STO files) are converted to XLS files, which can be easily manipulated and converted to Dataframe format using the Python Data Analysis Library (Pandas) \cite{Pandas}. Joint angles and torques were normalized in the range $[0,1]$ using min-max normalization as indicated by the following mathematical formula (Eq. \ref{eq:4}):

\begin{equation}
     x_{normalized} = \frac{x – \min{\{x_1,\dots x_T\}}}{\max{\{x_1,\dots x_T\}}-\min{\{x_1,\dots x_T\}}}
\label{eq:4}
\end{equation}

where $x$ is a sequence of either joint angle data $\mathbf{q}^N = \left\lbrace \mathbf{q}^N_1, ... , \mathbf{q}^N_T \right\rbrace$ or a sequence of joint torques $\mathbf{\tau}^N = \left\lbrace \mathbf{\tau}^N_1, ... , \mathbf{\tau}^N_T \right\rbrace$ for $t=1,...,T$ and $N=32$. Also, $min$ is the minimum and $max$ is the maximum value of the joint angles/torques sequences, respectively.

As mentioned in Section \ref{sec:dataset}, the Adapt Emotional Action dataset provides multiple trials for each motion, which have different durations, i.e. number of frames, ranging from $\sim 450-600$ frames. Thus, for every motion type, e.g. punching, we resample all motion trials to have the same number of frames using linear interpolation based on the trial with the longest duration. Our linear interpolation algorithm was developed using built-in functions from Python's SciPy Library \cite{SciPy}. The resampled motion trials will have a duration of approximately $600$ frames. 

\comm{
\begin{algorithm}
\caption{Resampling Time-Series Data Using Linear Interpolation}
\begin{algorithmic}[1]
\Require $joint\_angles$ (list of time-series data), $target\_length$ (desired resampling length)
\Ensure $trials$ (list of resampled trials)

\State Initialize empty lists: $synchronized\_trials$ and $trials$

\For{each $joint$ in $joint\_angles$}
    \For{each $trial$ in $joint$}
        \State Normalize original time points $t_o \gets$ Linspace of $[0,1]$ with length of $trial$
        \State Define new time points
        $t_t \gets$ Linspace of $[0,1]$ with $target\_length$
        \State $interpolator \gets$ Linear interpolation function for $(t_o, trial)$
        \State $resampled\_trial \gets$ Interpolated values at $t_t$
        \State Append $resampled\_trial$ to $synchronized\_trials$
    \EndFor
    \State Append a copy of $synchronized\_trials$ to $trials$
    \State Clear $synchronized\_trials$
\EndFor

\Return $trials$
\end{algorithmic}
\end{algorithm}}

\subsection{Training Scheme for Deep Learning Models}

All Deep Learning models constituting our Fatigue-PINN, namely Inverse and Forward Dynamics models, as well as 3CC-$\lambda$, are implemented and trained in Python Keras \cite{chollet2015keras}. The ID/FD BiLSTM models were trained supervisingly, based on training datasets containing the joint angles and torques produced from the preprocessing procedure outlined in the above sections of this Appendix. As discussed in section \ref{sec:methodology}, we opted for a joint-specific approach, meaning training multiple ID and FD models to produce predictions for each and every joint individually. In particular, joint kinematics from stacked motion trials, $\mathbf{q}^N$, are used as the training input of the ID model, while the torque actuating one joint (i.e. $N=1$), $\mathbf{\tau}$, from the respective multiple trials, is used as the ground-truth output of the model. On the contrary, for the FD model, joint torques from stacked motion trials$\mathbf{\tau}^N$ are utilized as the training input, and the angle of the same joint $\mathbf{q}$ is given as the ground-truth of the output, since the FD model performs the opposite task than ID, i.e. estimating joint angles based on joint torques. 

The ID and FD models were evaluated through a three-fold cross-validation approach applied across all motion trials of each motion type. Explanatory, the trials of each action are split into $3$ folds where the $\frac{2}{3}$ is being utilized as training data and the left-out $\frac{1}{3}$ as test data. The process is iteratively executed until each subset has been utilized as test data. Subsequently, the prediction accuracy of our model is computed by averaging the results across the three validation folds. Mean Square Error was utilized as the loss function of our ID and FD models. 

The 3CC-$\lambda$ was also trained supervisingly, utilizing datasets cited in \cite{frey-law2021}, containing fatigued joint torques measured after fatiguing experimental protocols. These datasets have very few time frames, and they are not subjected to the above pre-processing procedure, yet they only require normalization in the range of $[0, 1]$.  Equations (\ref{eq:3CC-loss1}), (\ref{eq:3CC-loss2}), and (\ref{eq:3CC-loss3}) describe the PINN-loss of our 3CC-$\lambda$ model.   

We trained our models with a batch size of $32$, for $1000$ epochs with early stopping, meaning that this number of epochs was never reached, avoiding over-fitting. Adam \cite{Kingma2014} was used as the optimizer with a learning rate of $0.001$.

Given that our 3CC-$\lambda$ model requires training based on datasets other than the Adapt Emotional dataset, it can be used as a pre-trained model. Therefore, the Fatigue-PINN framework can be either trained end-to-end or separately for each module. 

\subsection{Animation in Unity Environment}

To simulate our results in Unity Game Engine Environment \cite{Unity}, we may either: 
\begin{itemize}
    \item Re-implement our Fatigue-PINN framework in the Unity engine, using Numpy and Keras Python wrappers for C\#, the programming language for scripting in Unity. 
    \item Load our pre-trained models (i.e. initially train them in Python) in Unity using the same Python wrappers for C\#.
    \item Store the fatigued movement produced by our Fatigue-PINN, i.e. sequences of fatigued joint angles ($\acute{\hat{\mathbf{q}}}_t^N$), in a file format easily manipulated in C\#, e.g. a TXT file.   
\end{itemize}

The first two options involving the use of Python wrappers in C\# result in frame-drops (by $\sim10-15$ frames), since they require the initialization of the Python Engine as a background process, depleting the computational speed. Even though this may hinder the real-time execution of our Fatigue-PINN framework in Unity, our model can still be easily integrated into any motion synthesis pipeline operating on joint angles. The last option involves executing Fatigue-PINN as a standalone framework to produce fatigued motion sequences and subsequently loading the animation to a character.

The character utilized for showcasing our results in Unity is rigged. Character rigging is the 3D animation process where a set of manipulators and controls (e.g. joints, bones, etc.) are defined for a 3D virtual character. These controls constitute a "rig" or a digital skeleton, which aids the animator to correctly render the animation on the character \cite{Bhati2015}. However, the joints in the skeleton of the character as well as their orientations, do not match with those of the model used for musculoskeletal modeling. Therefore, we manually altered the number of joints as well as their orientations in the local coordinate system (the one of the respective joint) to be aligned with the OpenSim model. This correspondence is essential for accurately endowing the animation on the character, since the fatigued joint angles produced by Fatigue-PINN are translated in the OpenSim's local coordinate systems.

C\# scripting is utilized to animate the virtual character. This process entails configuring the 3D local joint angles of the character with the values of the fatigued joint kinematics produced by our Fatigue-PINN framework.  

\bibliographystyle{IEEEtran}
\bibliography{Bibliography_PINNs}

\begin{IEEEbiography}[{\includegraphics[width=1in,height=1.25in,clip,keepaspectratio]{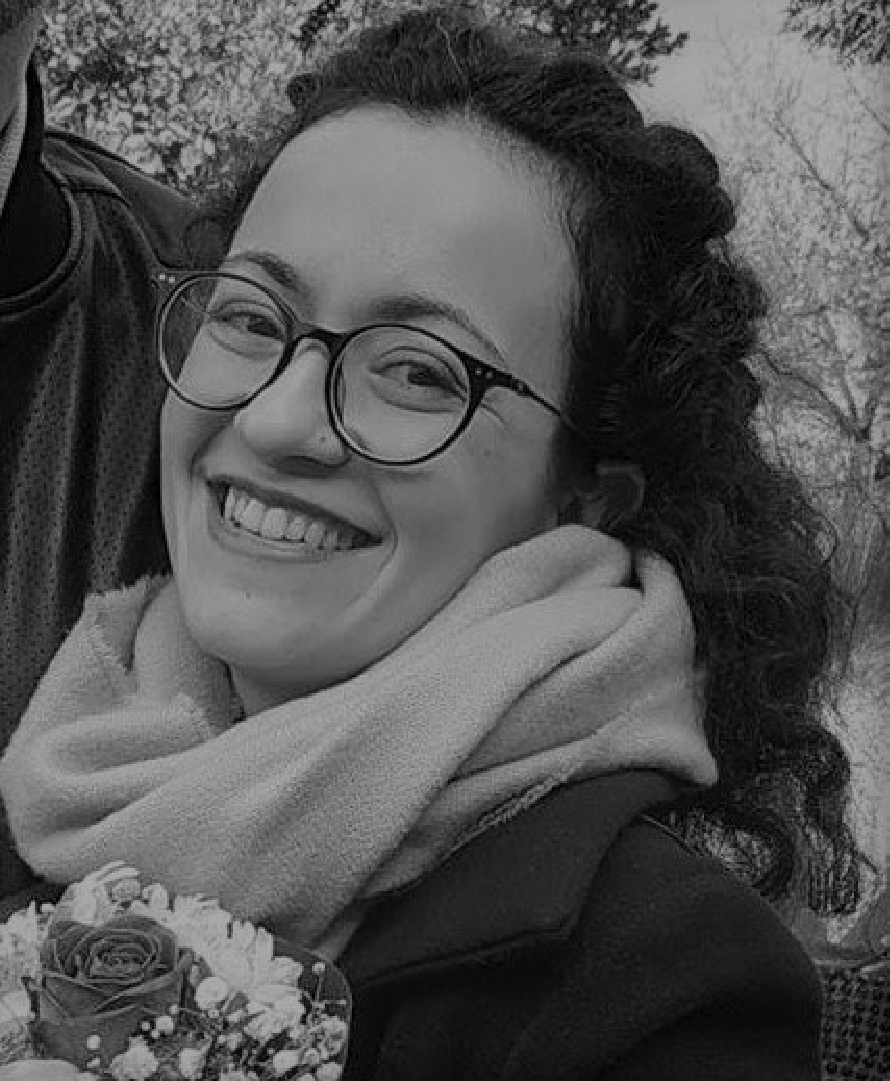}}]{Iliana Loi} Iliana Loi is a Ph.D. candidate at the Department of Electrical and Computer Engineering (ECE) of the University of Patras.  She obtained her Bachelor's degree from the Computer Engineering and Informatics Department of the University of Patras in 2018 and her
Master's in Biomedical Engineering in 2020. She is also a research assistant at the Visualization and Virtual Reality Group in the ECE since 2019 and has been working on research projects across various research fields including Artificial Intelligence, Computer Graphics, Biomedical Engineering, etc. Her research interests involve Machine/Deep Learning, Computer Animation, Musculoskeletal Modeling, and Biomechanical Simulation.
\end{IEEEbiography}

\begin{IEEEbiography}[{\includegraphics[width=1in,height=1.25in,clip,keepaspectratio]{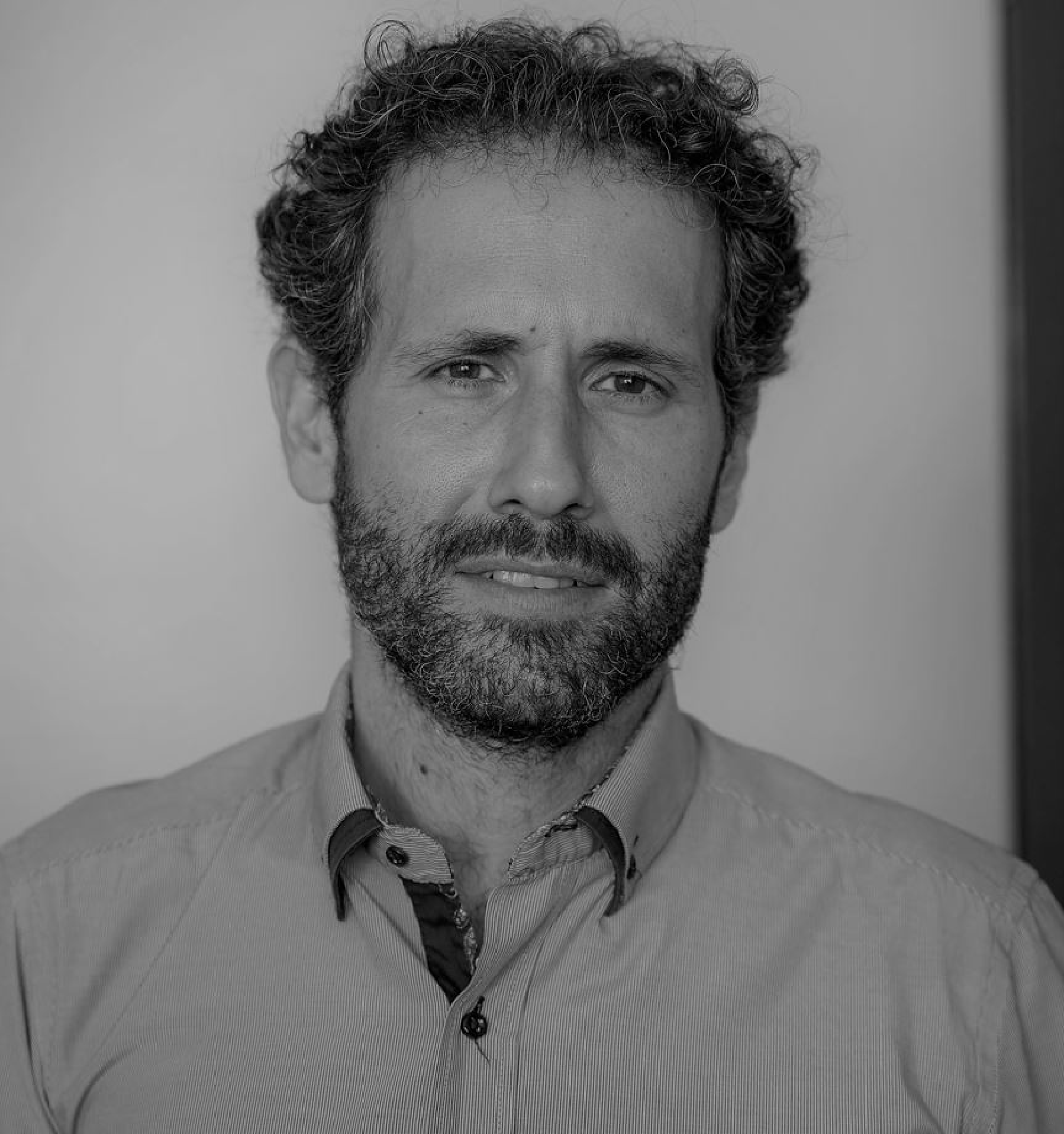}}]{Konstantinos Moustakas} Konstantinos Moustakas (Senior Member, IEEE) received the Diploma degree and the PhD in electrical and computer engineering from the Aristotle University of Thessaloniki, Greece, in 2003 and 2007 respectively. During 2007-2011 he served as a post-doctoral research fellow in the Information Technologies Institute, Centre for Research and Technology Hellas. He is currently a Professor at the Electrical and Computer Engineering Department of the University of Patras, Head of the Visualization and Virtual Reality Group, Director of the Wire Communications and Information Technology Laboratory and Director of the MSc Program on Biomedical Engineering of the University of Patras. He serves as an Academic Research Fellow for ISI/Athena research center. His main research interests include virtual, augmented and mixed reality, 3D geometry processing, haptics, virtual physiological human modeling, information visualization, physics-based simulations, computational geometry, computer vision. During the latest years, he has been the (co)author of more than 250 papers in refereed journals, edited books, and international conferences. His research work has received several awards. He has participated in more than 30 research and development projects funded by the EC and the Greek Secretariat of Research and Technology, while he has served as the coordinator or scientific coordinator in 4 of them. He has also been a member of the organizing committee of several international conferences. He is a senior member of the IEEE, the IEEE Computer Society and member of Eurographics.
\end{IEEEbiography}



\EOD

\end{document}